\documentclass[12pt, draftclsnofoot, onecolumn]{IEEEtran}
\usepackage{setspace}
\usepackage[dvips]{graphicx}
\usepackage{epstopdf}
\usepackage{amsmath, amsthm, amssymb, amsfonts, mathtools}
\usepackage{cases, multirow}
\usepackage{srcltx,color}
\usepackage{cite}
\usepackage{courier}
\usepackage{subfigure}
\usepackage{float}
\long\def\comment#1{}

\newtheorem{prop}{Proposition}

\def\figref#1{Fig.~\ref{#1}}
\def\be{\begin{equation} }
\def\ee{\end{equation} }

\title{Secrecy Capacity Analysis over $\kappa-\mu$ Fading Channels: Theory and Applications}

\begin{document}
  
\author{
Nidhi Bhargav, Simon L. Cotton and David E. Simmons
\IEEEcompsocitemizethanks{
  \IEEEcompsocthanksitem 
  N. Bhargav and S. L. Cotton are with the Wireless Communications Laboratory, Institute of Electronics, Communications and Information Technology, The Queen's University of Belfast, Queen's Road, Belfast, BT3 9DT, UK. Email: \{nbhargav01, simon.cotton\}@qub.ac.uk.
	
  D. E. Simmons is with the Department of Engineering Science, University of Oxford, Parks Road, Oxford, OX1 3PJ, UK. Email: david.simmons@eng.ox.ac.uk.
	
	This work has been submitted to the IEEE for possible publication. Copyright may be transferred without notice, after
which this version may no longer be accessible.

  }
}

\maketitle

% %%% Line Spacing Control %%%
% % \singlespacing
% % \onehalfspacing
%\doublespacing
% % \setstretch{2.0}

\begin{abstract}
\begin{spacing}{1}
In this paper, we consider the transmission of confidential information over a $\kappa$-$\mu$ fading channel in the presence of an eavesdropper, who also observes $\kappa$-$\mu$ fading. In particular, we obtain novel analytical solutions for the probability of strictly positive secrecy capacity (SPSC) and the lower bound of secure outage probability (SOP$^L$) for channel coefficients that are positive, real, independent and non-identically distributed ($i.n.i.d.$). We also provide a closed-form expression for the probability of SPSC when the $\mu$ parameter is assumed to only take positive integer values. We then apply the derived results to assess the secrecy performance of the system in terms of the average signal-to-noise ratio (SNR) as a function of the $\kappa$ and $\mu$ fading parameters. We observed that for fixed values of the eavesdropper's average SNR, increases in the average SNR of the main channel produce a higher probability of SPSC and a lower secure outage probability (SOP). It was also found that when the main channel experiences a higher average SNR than the eavesdropper's channel, the probability of SPSC improved while the SOP was found to decrease with increasing values of $\kappa$ and $\mu$ for the legitimate channel. The versatility of the $\kappa$-$\mu$ fading model, means that the results presented in this paper can be used to determine the probability of SPSC and SOP$^L$ for a large number of other fading scenarios such as Rayleigh, Rice (Nakagami-$n$), Nakagami-$m$, One-Sided Gaussian and mixtures of these common fading models. Additionally, due to the duality of the analysis of secrecy capacity and co-channel interference, the results presented here will also have immediate applicability in the analysis of outage probability in wireless systems affected by co-channel interference and background noise. To demonstrate the efficacy of the novel formulations proposed here, we use the derived equations to provide a useful insight into the probability of SPSC for a range of emerging applications such as cellular device-to-device, vehicle-to-vehicle and body centric fading channels using data obtained from field measurements. 
\end{spacing}
\end{abstract}

\begin{IEEEkeywords}
Fading channels, $\kappa-\mu$ fading, secrecy capacity, co-channel interference, device-to-device communications, vehicular communications, body centric communications.
\end{IEEEkeywords}

%%%%%%%%%%%%%%%%%%%%%%%%%%%%%%%%%%%%%%%%%%%%%%%%%%%%%%%%%%%
\section{Introduction}
%%%%%%%%%%%%%%%%%%%%%%%%%%%%%%%%%%%%%%%%%%%%%%%%%%%%%%%%%%%

With the proliferation of smart devices, driven by applications such as the internet of things (IoT)~\cite{IoT}, device-to-device communications (D2D)~\cite{cotton2015human} and wearable sensors~\cite{cotton2009experimental,cotton2007characterization,cotton2006indoor}, privacy and security in wireless networking systems have once again been brought to the forefront. The wireless medium utilized by each of these applications has an inherent broadcast nature that makes it particularly susceptible to eavesdropping. Traditionally, these systems have attained secure communications by employing classical cryptographic techniques; e.g., RSA or AES~\cite{rivest1978method}. Unfortunately, these algorithms are entirely disjoint from the physical nature of the wireless medium as they assume that the physical layer provides an error-free link. More recently, there has been growing interest in information-theoretic security that exploits the random nature of the wireless channel to guarantee the confidential transmission of messages~\cite{maurer1993secret}. It is widely believed that using this type of approach will provide the strictest form of security for physical layer communications~\cite{bloch2008wireless}.

The notion of perfect information-theoretic secrecy, i.e. $I(M;C)=0$, where $I(\cdot;\cdot)$ denotes mutual information, $M$ is the plane text message and $C$ is its corresponding encryption, was first presented by Shannon~\cite{shannon1949communication}. These ideas were later developed by Wyner~\cite{wyner1975wire}, in which he introduced the wiretap channel. Under the assumption that the wiretapper's channel is a probabilistically degraded version of the main channel, he studied the trade-off between the information rate and the achievable secrecy level for a wiretap channel and showed that it is possible to achieve a non-zero secrecy capacity. The secrecy capacity is defined as the largest transmission rate from the source to the destination, at which the eavesdropper is unable to obtain any information. Csisz\'{a}r and K\"{o}rner~\cite{csiszar1978broadcast} later extended Wyner's work to non-degraded channels where the source has a common message for both the intended receiver and the eavesdropper in addition to the confidential information for the intended recipient. Further developments were made in~\cite{leung1978gaussian}, where it was shown that it is possible to achieve secure communication in the presence of an eavesdropper over an additive white Gaussian noise (AWGN) channel provided the channel capacity of the legitimate user was greater than the eavesdropper's. The secrecy capacity was then shown to be equal to the difference between the two channel capacities. 

The effect of fading on secrecy capacity was studied in~\cite{li2010secrecy,liang2007secrecy,gopala2008secrecy,tekin2008gaussian,liang2008secure,barros2006secrecy,parada2005secrecy,wang2007secrecy,nak5469979multi,nak5501963infotheory,secrecylognormal,secrecyweibull,6338984}.  Li \textit{et al.}~\cite{li2010secrecy}, Liang \textit{et al.}~\cite{liang2007secrecy} and Gopal \textit{et al.}~\cite{gopala2008secrecy} characterized the secrecy capacity of ergodic fading channels and presented power and rate allocation schemes for secure communication. In~\cite{tekin2008gaussian} and~\cite{liang2008secure}, the secrecy capacity for multiple access and broadcast channels was considered. Barros and Rodrigues~\cite{barros2006secrecy} showed that with signal fluctuation due to fading, information-theoretic security is achievable even when the eavesdropper's channel is of better average quality than that of the intended recipient. They analyzed the SOP and the outage secrecy capacity for Rayleigh fading channels when both the transmitter and the receiver are equipped with a single antenna in the presence of a solitary eavesdropping party.  A similar analysis for a system consisting of a single antenna at the transmitter and multiple antennas at the receiver was presented in~\cite{parada2005secrecy}.  This was extended to multiple eavesdropping parties in~\cite{wang2007secrecy} and over Nakagami-$m$ fading channels in~\cite{nak5469979multi} and~\cite{nak5501963infotheory}. It was found that the SOP increases with the number of eavesdroppers and the average SNR of the eavesdropper. The outage secrecy capacity was found to increase with the Nakagami-$m$ parameter, since an increase in $m$ decreases the severity of fading in the channel. More recently, the secrecy characteristics of other commonly encountered fading models such as lognormal, Weibull and Rice have also been studied. For example, the probability of SPSC of lognormal fading channels was studied in~\cite{secrecylognormal}, while the probability of SPSC and SOP$^L$ of Weibull fading channels was investigated in~\cite{secrecyweibull}. In~\cite{6338984} a secrecy capacity analysis over Rice/Rice fading channels was conducted and the probability of non-zero secrecy capacity was determined. 

While a number of important performance measures for $\kappa$-$\mu$ fading channels~\cite{4231253} have previously been developed such as energy detection based spectrum sensing~\cite{6359882,annamalai2011unified,6803901} and outage probability analysis in interference-limited scenarios with restricted values of $\mu$~\cite{6803901}, to the best of the author's knowledge the secrecy capacity of $\kappa$-$\mu$ fading channels has yet to be reported in the open literature. Due to the equivalency pointed out in~\cite{7001255}, the results presented here will also have immediate applicability in the analysis of outage probability in cellular systems affected by co-channel interference and background noise, and the calculation of outage probability in interference-limited scenarios. Indeed the new equations proposed here provide an alternative, more general result than those presented in~\cite{6803901} (wherein the outage probability in interference-limited scenarios is restricted to particular values of the $\mu$ parameter) and allow the calculation of the relevant capacity formulations for arbitrary, real, positive values of the $\mu$ parameter. Motivated by this, we analyze the secrecy capacity of $\kappa$-$\mu$ fading channels in which we assume the eavesdropper to be passive and the channel state information (CSI) of the eavesdropper and the intended recipient are not available at the transmitter. 

The main contributions of this paper are now listed as follows. Firstly, we derive novel analytical expressions for the probability of SPSC and SOP$^L$ over $\kappa$-$\mu$ fading channels for coefficients that are positive, real, \textit{i.n.i.d.} for both the legitimate and non-legitimate parties. Secondly, we provide an exact closed-form solution for the probability of SPSC over $\kappa$-$\mu$ fading channels when the $\mu$ parameter is assumed to take positive integer values. These expressions have been subsequently verified by reduction to known special cases. Thirdly and most importantly, because the $\kappa$-$\mu$ fading model~\cite{4231253} contains a number of other well-known fading models as special cases, the novel formulations presented in this paper unify the secrecy capacity of Rayleigh, Rice (Nakagami-$n$), Nakagami-$m$ and One-Sided Gaussian fading channels, and their mixtures. Therefore they can be used to provide a useful insight into the secrecy capacity of eavesdropping scenarios which undergo generalized fading conditions. Finally, we provide an important application of these results to estimate the probability of SPSC of a number of emerging wireless applications such as cellular device-to-device, vehicle-to-vehicle and body centric communications using data obtained from real channel measurements.  

The remainder of this paper is organized as follows. Section II provides a brief overview of the $\kappa$-$\mu$ fading model. Section III explains the system model while Section IV provides the derivation of novel analytical and closed form expressions for the probability of SPSC and SOP$^L$. Section V provides an overview of the model parameters required to obtain the secrecy capacity of the common fading models derived from the $\kappa$-$\mu$ fading model; this is followed by some numerical results. Section VI discusses some of the applications of this paper. Lastly, Section VII finishes the paper with some concluding remarks.

%%%%%%%%%%%%%%%%%%%%%%%%%%%%%%%%%%%%%%%%%%%%%%%%%%%%%%%%%%%
\section{An Overview of the $\kappa$-$\mu$  Fading Model} 
%%%%%%%%%%%%%%%%%%%%%%%%%%%%%%%%%%%%%%%%%%%%%%%%%%%%%%%%%%%
The $\kappa$-$\mu$ fading model was originally conceived for modeling the small-scale variations of a fading signal under line-of-sight (LOS) conditions in non-homogeneous environments~\cite{4231253}. The $\kappa$-$\mu$  fading signal is a composition of clusters of multipath waves with scattered waves of identical power with a dominant component of arbitrary power found within each cluster. The received signal envelope, $R$, of the $\kappa$-$\mu$ fading model may be expressed in terms of the in-phase and quadrature components of the fading signal such that~\cite[eq. 6]{4231253}
\begin{equation}
{R^2} = \mathop \sum \limits_{i = 1}^\mu  {\left( {{X_i} + {p_i}} \right)^2} + \mathop \sum \limits_{i = 1}^\mu  {\left( {{Y_i} 
+ {q_i}} \right)^2}\label{eq:1}
\end{equation}
where $\mu$ is the number of multipath clusters, $X_i$ and $Y_i$ are mutually independent Gaussian random processes with mean $E[{X_i}] = E[{Y_i}] = 0$ and variance $E\left[ {{X_i}^2} \right] = E\left[ {{Y_i}^2} \right] = {\sigma ^2}$ (i.e., the power of the scattered waves in each clusters). Here $p_i$ and $q_i$ are the mean values of the in-phase and quadrature phase components of multipath cluster $i$ and ${d^2} = \mathop \sum \limits_{i = 1}^\mu  {p_i}^2 + {q_i}^2.$ Letting $\gamma$ represent the instantaneous signal-to-noise-ratio (SNR) of a $\kappa$-$\mu$ fading channel, then its probability density function (PDF), ${f_\gamma }\left( \gamma  \right)$, is obtained from the envelope PDF  given in~\cite[eq. 11]{4231253} via a transformation of variables $\left( {r = \sqrt {\gamma { \ {\hat r}^2}/\bar \gamma } } \right)$ as, 
\begin{equation}
{f_\gamma }\left( \gamma  \right) = \frac{{\mu {{\left( {1 + \kappa } \right)}^{\frac{{\mu  + 1}}{2}}}{\gamma ^{\frac{{\mu  - 1}}{2}}}{e^{\frac{{ - \mu \left( {1 + \kappa } \right)\gamma }}{{\bar \gamma }}}}}}{{{\kappa ^{\frac{{\mu  - 1}}{2}}}{{\bar \gamma }^{\frac{{\mu  + 1}}{2}}}{e^{\mu \kappa }}}}{I_{\mu  - 1}}\left( {2\mu \sqrt {\frac{{\kappa \left( {1 + \kappa } \right)\gamma }}{{\bar \gamma }}} } \right)\label{eq:2}
\end{equation}
where $\kappa>0$ is the ratio of the total power of the dominant components $(d^2)$ to that of the scattered waves $(2\mu\sigma^2)$ in each of the clusters, $\mu>0$ is related to the number of multipath clusters and is given by $\mu  = \frac{{{\mathbb{E}^2}\left( \gamma  \right)\left( {1 + 2\kappa } \right)}}{{\mathbb{V} \left( \gamma  \right){{\left( {1 + \kappa } \right)}^2}}}$ where  $\mathbb{E(\cdot)}$ and $\mathbb{V(\cdot)}$ denote the expectation and variance operators, respectively, $\bar \gamma  = \mathbb{E}\left( \gamma  \right)$, is the average SNR and ${I_n}(\cdot)$ is the modified Bessel function of the first kind and order $n$. As indicated in~\cite{4231253}, it should be noted that the $\mu$ parameter can take non-integer values which can be due to the non-zero correlation between the in-phase and quadrature components of each cluster, non-zero correlation between the multipath clusters or non-Gaussian nature of the in-phase and quadrature components. The cumulative distribution function (CDF) of $\gamma$ can be obtained from~\cite[eq. 3]{4231253} as,
\begin{equation}
{F_\gamma }\left( \gamma  \right) = 1 - {Q_\mu }\left[ {\sqrt {2\kappa \mu } ,\sqrt {\frac{{2\left( {1 + \kappa } \right)\mu \gamma }}{{\bar \gamma }}} } \right]\label{eq:3}
\end{equation}
where ${Q_ \cdot }\left( { \cdot , \cdot } \right)$ is the generalized Marcum $Q$-function defined in~\cite[eq. 4.60]{Digitalcomm} as, 
\begin{equation}
{Q_M}\left( {\alpha ,\beta } \right) = \frac{1}{{{\alpha ^{M - 1}}}}\mathop \smallint \limits_\beta ^\infty  {x^M}{e^{ - \left( {\frac{{{x^2} + {\alpha ^2}}}{2}} \right)}}{I_{M - 1}}\left( {\alpha x} \right)dx.\label{eq:4}
\end{equation}
The $\kappa$-$\mu$ distribution is a generalized fading model which also contains as special cases other important distributions such as the Rice ($\mu$ = 1 and $\kappa$ = $K$), Nakagami-$m$ ($\kappa$ $\rightarrow$ 0 and $\mu$ = $m$), Rayleigh ($\mu$ = 1 and $\kappa$ $\rightarrow$ 0) and One-Sided Gaussian ($\mu$ = 0.5 and $\kappa$ $\rightarrow$ 0).

%%%%%%%%%%%%%%%%%%%%%%%%%%%%%%%%%%%%%%%%%%%%%%%%%%%%%%%%%%%
\section{The System Model }
\begin{figure}[htbp]
	\centering
	 \includegraphics[width=0.55\textwidth]{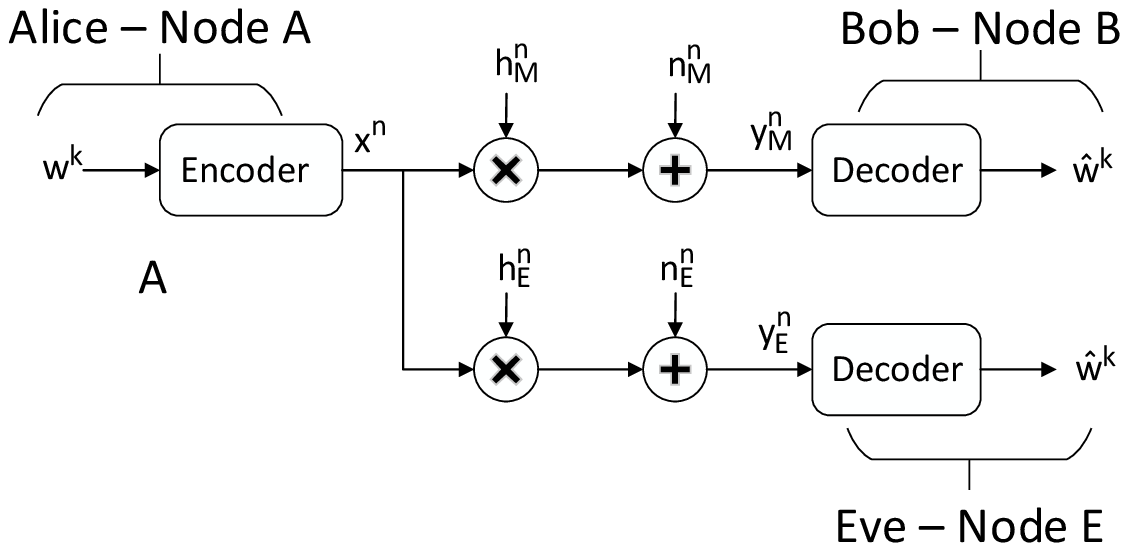}
	 \caption{The proposed system model.}
	\label{fig:img-1}
\end{figure}

%%%%%%%%%%%%%%%%%%%%%%%%%%%%%%%%%%%%%%%%%%%%%%%%%%%%%%%%%%%
Consider the system model of secure data transmission shown in \figref{fig:img-1}. The legitimate transmitter, Alice (node A), wishes to communicate secretly with the legitimate receiver, Bob (node B), while a third party, Eve (node E), is attempting to eavesdrop. We assume that the main and eavesdropper's channels both experience $\kappa$-$\mu$ fading. Alice wishes to send a message, $w^k= [w(1),w(2),w(3) \ldots w(k)]$ to Bob. At the transmitter, the message $w^k$ is encoded into a codeword, $x^n= [x(1),x(2),x(3) \ldots x(n)]$, for transmission over the channel. The signal received by Bob and Eve can be written as,  
\begin{align}
{y_M}\left( i \right) = {h_M}\left( i \right)x\left( i \right) + {n_M}\left( i \right)
\\ {y_E}\left( i \right) = {h_E}\left( i \right)x\left( i \right) + {n_E}\left( i \right)
\end{align}
where $h_M (i)$ and $h_E (i)$ are the quasi-static $\kappa$-$\mu$ fading coefficients of the main and the eavesdropper's channels respectively (i.e., $h_M (i)= h_M  \forall i$ and $h_E (i)= h_E  \forall i$) and $n_M (i)$ and $n_E (i)$ are the zero-mean circularly symmetric complex Gaussian noise random variables with unit variance at Bob and Eve respectively.
%\begin{figure*}[!t]
     %\centering 
   
We assume that the fading coefficients of Bob and Eve's channels, although random, are constant during the transmission of an entire codeword and independent of each other. Letting $P$, $N_M$ and $N_E$ represent the average transmit power, noise power in the main channel, and noise power in the eavesdropper's channel respectively, then, the corresponding instantaneous SNR's at the legitimate receiver and the eavesdropper are given by, $\gamma_M = \frac{P |h_M |^2}{N_M}$  and $\gamma_E= \frac{P |h_E |^2}{N_E}$ , while the average SNR's are given by, $\bar\gamma_M= \frac{P \mathbb{E}(|h_M |^2 )}{N_M}$ and  $\bar\gamma_E=\frac{P \mathbb{E}(|h_E |^2 )}{N_E}$.
Now let us consider the channel components of Bob and Eve which are assumed to be\textit{ i.n.i.d.} random variables with parameters \{$\kappa_M$, $\mu_M$, $\bar\gamma_M$\} and \{$\kappa_E$, $\mu_E$, $\bar\gamma_E$\}, respectively. The PDF's of $\gamma_M$  and $\gamma_E$ can be re-written from $\eqref{eq:2}$  as,
\begin{equation}   
{f_{{\gamma _M}}}\left( {{\gamma _M}} \right) = \frac{{{\mu _M}{{\left( {1 + {\kappa _M}} \right)}^{\frac{{{\mu _M} + 1}}{2}}}{\gamma _M}^{\frac{{{\mu _M} - 1}}{2}}{e^{\frac{{ - {\mu _M}\left( {1 + {\kappa _M}} \right){\gamma _M}}}{{\overline {{\gamma _M}} }}}}}}{{{\kappa _M}^{\frac{{{\mu _M} - 1}}{2}}{{\overline {{\gamma _M}} }^{\frac{{{\mu _M} + 1}}{2}}}{e^{{\mu _M}{\kappa _M}}}}}{I_{{\mu _M} - 1}}\left( {2{\mu _M}\sqrt {\frac{{{\kappa _M}\left( {1 + {\kappa _M}} \right){\gamma _M}}}{{\overline {{\gamma _M}} }}} } \right)\label{eq:5}
\end{equation}
\begin{equation}   
{f_{{\gamma _E}}}\left( {{\gamma _E}} \right) = \frac{{{\mu _E}{{\left( {1 + {\kappa _E}} \right)}^{\frac{{{\mu _E} + 1}}{2}}}{\gamma _E}^{\frac{{{\mu _E} - 1}}{2}}{e^{\frac{{ - {\mu _E}\left( {1 + {\kappa _E}} \right){\gamma _E}}}{{\overline {{\gamma _E}} }}}}}}{{{\kappa _E}^{\frac{{{\mu _E} - 1}}{2}}{{\overline {{\gamma _E}} }^{\frac{{{\mu _E} + 1}}{2}}}{e^{{\mu _E}{\kappa _E}}}}}{I_{{\mu _E} - 1}}\left( {2{\mu _E}\sqrt {\frac{{{\kappa _E}\left( {1 + {\kappa _E}} \right){\gamma _E}}}{{\overline {{\gamma _E}} }}} } \right).\label{eq:6}
\end{equation}
Likewise, the CDF's of $\gamma_M$  and $\gamma_E$ can be re-written from $\eqref{eq:3}$ as,
\begin{equation}
{F_{{\gamma _M}}}\left( {{\gamma _M}} \right) = 1 - {Q_{{\mu _M}}}\left[ {\sqrt {2{\kappa _M}{\mu _M}} ,\sqrt {\frac{{2\left( {1 + {\kappa _M}} \right){\mu _M}{\gamma _M}}}{{\overline {{\gamma _M}} }}} } \right]\label{eq:7}
\end{equation}
\begin{equation}
{F_{{\gamma _E}}}\left( {{\gamma _E}} \right) = 1 - {Q_{{\mu _E}}}\left[ {\sqrt {2{\kappa _E}{\mu _E}} ,\sqrt {\frac{{2\left( {1 + {\kappa _E}} \right){\mu _E}{\gamma _E}}}{{\overline {{\gamma _E}} }}} } \right].\label{eq:8}
\end{equation}

%%%%%%%%%%%%%%%%%%%%%%%%%%%%%%%%%%%%%%%%%%%%%%%%%%%%%%%%%%%
\section{Secrecy Capacity in $\kappa$-$\mu$ Fading Channels} 
%%%%%%%%%%%%%%%%%%%%%%%%%%%%%%%%%%%%%%%%%%%%%%%%%%%%%%%%%%%

In this section, we derive an analytical expression for the SOP$^L$. We also arrive at both an analytical and closed form expression for the probability of SPSC over $\kappa$-$\mu$ fading channels. Herein, it should be noted that we adopt the notation {$M / E$} to describe the fading conditions experienced by the main channel ($M$), i.e. the intended recipient, and the eavesdropper ($E$). For example $\kappa$-$\mu$ / $\kappa$-$\mu$ indicates that the main and eavesdropper's channels are both subject to\textit{ i.n.i.d.} $\kappa$-$\mu$ fading. Of course, because of the generality of the $\kappa$-$\mu$ fading model, $M$ and $E$ can be readily interchanged with the Rayleigh, Nakagami-$m$, Rice (Nakagami-$n$) and One-sided Gaussian models which all appear as special cases of this fading model.

In passive eavesdropping scenarios, where the CSI of the eavesdropper and the intended recipient are not available at the transmitter, perfect secrecy is not guaranteed. Hence, we adopt the probability of non-zero secrecy capacity and SOP as useful secrecy metrics to characterize the system performance. For the system under consideration, the capacity of the main channel is given by $C_M = \log_2 \left( 1   +   \gamma_M \right)$ and the capacity of the eavesdropper's channel is given by $C_E = \log_2 \left( 1   +   \gamma_E \right)$. From~\cite{bloch2008wireless}, we define the secrecy capacity, $C_S$, for one realization of the SNR pair ($\gamma_M$, $\gamma_E$) of the quasi-static complex fading wiretap channel as,
\begin{equation}
{C_S} = \left\{ {\begin{array}{*{20}{c}}
{{{\log }_2}\left( {1 + {\gamma _M}} \right) - {{\log }_2}\left( {1 + {\gamma _E}} \right),\left( {{\gamma _M} > {\gamma _E}} \right)}\\
{0~~~~~~~~~~~~~~~~~~~~~~~~~~~~~~~~~~~,\left( {{\gamma _M} \le {\gamma _E}} \right)}
\end{array}} \right.
\end{equation}

\subsection{SOP Analysis}

The outage probability of secrecy capacity is the probability that the instantaneous secrecy capacity falls below a target secrecy rate  $R_S$ $(R_S \ge 0)$ and is defined in~\cite{bloch2008wireless} as,   
\begin{equation}
{P_{out}}\left( {{R_S}} \right) = {P_r}\left( {{C_S} \le {R_S}} \right).\label{eq:9}
\end{equation}
Performing the necessary mathematical manipulations, we obtain,	
\begin{equation}
{P_{out}}\left( {{R_S}} \right) = {P_r}\left[ {{\gamma _M} \le {e^{R_S} }\left( {1 + {\gamma _E}} \right) - 1} \right] \label{eq:1a}
\end{equation}
which can then be expressed as,
\begin{multline}
{P_{out}}\left( {{R_S}} \right)  =  \frac{{{\mu _E}{{\left( {1 + {\kappa _E}} \right)}^{\frac{{{\mu _E} + 1}}
{2}}}}}{{{\kappa _E}^{\frac{{{\mu _E} - 1}}{2}}{{\overline {{\gamma _E}} }^{\frac{{{\mu _E} + 1}}
{2}}}{e^{{\mu _E}{\kappa _E}}}}}
\intop_0^\infty  {{\gamma _E}^{\frac{{{\mu _E} - 1}}{2}}
e^{\frac{{ - {\mu _E}\left( {1 + {\kappa _E}} \right){\gamma _E}}}{{\overline {{\gamma _E}} }}}}{I_{{\mu _E} - 1}}\left( {2{\mu _E}\sqrt {\frac{{{\kappa _E}\left( {1 + {\kappa _E}} \right)
{\gamma _E}}}{{\overline {{\gamma _E}} }}} } \right)\\ \times \left(1 - {Q_{{\mu _M}}}\left( {\sqrt 
{2{\kappa _M}{\mu _M}} ,\sqrt {\frac{{2\left( {1 + {\kappa _M}} \right)\left( {{e^{R_S} }\left( 
{1 + {\gamma _E}} \right) - 1} \right){\mu _M}}}{{\overline {{\gamma _M}} }}} } \right)\right) d{\gamma _E}.
 \label{eq:1b}\end{multline}
\begin{IEEEproof}
 See Appendix \ref{app:A}.
\end{IEEEproof}

At present, due to the complicated form of the integral contained in $\eqref{eq:1b}$ it is not possible to obtain a closed-form expression for the SOP, therefore, we derive the lower bound of SOP as follows~\cite{secrecyweibull},

\begin{equation}\begin{aligned}
{P_{out}}\left( {{R_S}} \right) &= {P_r}\left[ {{\gamma _M} \le {e^{R_S} }\left( {1 + {\gamma _E}} \right) - 1} \right] \\ &\geq {\mathrm{SOP}^L} = {P_r}\left[ {\gamma _M} \le {e^{R_S}}{\gamma _E}\right].
 \label{eq:1c}\end{aligned}\end{equation}

Consider the Marcum-Q function in $\eqref{eq:1b}$. This can be re-written as,
\begin{align}
{Q_{{\mu _M}}}\left( {\sqrt {2{\kappa _M}{\mu _M}} ,\sqrt {\frac{{2\left( {1 + {\kappa _M}} \right){\mu _M}{e^{{R_S}}}{\gamma _E}}}{{\overline {{\gamma _M}} }}\left( {1 + \frac{{{e^{{R_S}}} - 1}}{{{e^{{R_S}}}{\gamma _E}}}} \right)} } \right){\gamma _E}\mathop  \to \limits_ =  \infty \nonumber \\
{Q_{{\mu _M}}}\left( {\sqrt {2{\kappa _M}{\mu _M}} ,\sqrt {\frac{{2\left( {1 + {\kappa _M}} \right){\mu _M}{e^{{R_S}}}{\gamma _E}}}{{\overline {{\gamma _M}} }}} } \right).
\label{eq:1z}\end{align}

Now, using $\eqref{eq:6}$, $\eqref{eq:7}$ and $\eqref{eq:1z}$ the lower bound of SOP is obtained as,
\begin{multline}
{\mathrm{SOP}^L} =  \frac{{{\mu _E}{{\left( {1 + {\kappa _E}} \right)}^{\frac{{{\mu _E} + 1}}
{2}}}}}{{{\kappa _E}^{\frac{{{\mu _E} - 1}}{2}}{{\overline {{\gamma _E}} }^{\frac{{{\mu _E} + 1}}
{2}}}{e^{{\mu _E}{\kappa _E}}}}}
\intop_0^\infty  {{\gamma _E}^{\frac{{{\mu _E} - 1}}{2}}
e^{\frac{{ - {\mu _E}\left( {1 + {\kappa _E}} \right){\gamma _E}}}{{\overline {{\gamma _E}} }}}}{I_{{\mu _E} - 1}}\left( {2{\mu _E}\sqrt {\frac{{{\kappa _E}\left( {1 + {\kappa _E}} \right)
{\gamma _E}}}{{\overline {{\gamma _E}} }}} } \right)\\ \times \left(1 - {Q_{{\mu _M}}}\left( {\sqrt 
{2{\kappa _M}{\mu _M}} ,\sqrt {\frac{{2\left( {1 + {\kappa _M}} \right)\left( {{e^{R_S} } 
{{\gamma _E}}} \right){\mu _M}}}{{\overline {{\gamma _M}} }}} } \right)\right) d{\gamma _E}.
 \label{eq:1d}\end{multline}

The solution for $\eqref{eq:1d}$ can be obtained via the following proposition.
\begin{prop} For the arbitrary real and positive  $\mu_M$, $\mu_E$, $\kappa_M$ and $\kappa_E$ $\eqref{eq:1d}$ can be evaluated as, 
\begin{align}
{\mathrm{SOP}^L} &= \frac{{{{{{\beta _E}}}^{\frac{{{\mu _E} + 1}}{2}}}}}{{{{{\alpha_E}}^{\frac{{{\mu _E} - 1}}{2}}}{e^{{\alpha _E}}}}}\sum\limits_{k = 0}^\infty  {\frac{{{{\left( {\sqrt {{\alpha_E}{\beta _E}} } \right)}^{{\mu _E}- 1+2k}}{\rm{\Gamma }}\left( {{\mu _E}+k} \right)}}{{k!{\rm{\Gamma }}\left( {{\mu _E}+k} \right)}}} {\left( {\frac{1}{{{\beta _E}}}} \right)^{{\mu _E}+k}}-\frac{{{{{{\beta _E}}}^{\frac{{{\mu _E} + 1}}{2}}}}}{{{{{\alpha _E}}^{\frac{{{\mu _E} - 1}}{2}}}{e^{{\alpha _E}}}}}\nonumber \\
&\times \sum\limits_{k = 0}^\infty  {\sum\limits_{l = 0}^\infty  {\frac{{{{\left( {\sqrt {{\alpha_E}{\beta _E}} } \right)}^{{\mu _E} - 1 + 2k}}{{\left( {2{\alpha _M}} \right)}^l}{\left({\beta _M}{e^{R_S}}\right)}^{{\mu _M} + l}{\rm{\Gamma }}\left( {{\mu _E} + k + {\mu _M} + l} \right)}}{{k!{\rm{\Gamma }}\left( {{\mu _E} + k} \right)l!{\rm{\Gamma }}\left( {{\mu _M} + l} \right){2^{l}}{e^{{\alpha _M}}}\left( {{\mu _E} + k} \right){{\left( {{\beta _M {e^{R_S} }} + {\beta _E}} \right)}^{{\mu _E} + k + {\mu _M} + l}}}}} } \nonumber \\
&\times {_2}{F_1}\left( {1,{\mu _E} + k + {\mu _M} + l;{\mu _E} + k + 1;\frac{{{\beta _E}}}{{{\beta _M {e^{R_S} }} + {\beta _E}}}} \right)
\label{eq:1e}\end{align}
where $a=  1/\bar\gamma_M$,  $b=  1/\bar\gamma_E$, ${\beta _M} = ({\kappa _M} + 1)a{\mu _M}$, ${\beta _E} = ({\kappa _E} + 1)b{\mu _E}$, ${\alpha _M} = {\kappa _M}{\mu _M}$, ${\alpha _E} = {\kappa _E}{\mu _E}$ and $\mathstrut_2 F_1(\cdot\ ,\cdot \ ;\ \cdot  \ ;\ \cdot)$  is the Gauss hypergeometric function~\cite{TofI}. \label{prop:1a}
\end{prop}
\begin{IEEEproof}
 See Appendix \ref{app:B}.
\end{IEEEproof}

\subsection{SPSC Analysis}
In this subsection we examine the condition for the existence of strictly positive secrecy capacity. This occurs
as a special case of the secrecy outage probability when the target secrecy rate, $R_S = 0$. According to~\cite{bloch2008wireless} the probability of non zero secrecy capacity is defined as,
\begin{equation}
{P_0}={P_r}\left( {{C_S} > {0}} \right) = {P_r}\left( {{\gamma_M} > {\gamma_E}} \right) 
\label{eq:1f}\end{equation}
Performing the necessary mathematical manipulations, the probability of SPSC can be expressed as,
\begin{multline}
{P_\alpha } = \frac{{{\mu _E}{{\left( {1 + {\kappa _E}} \right)}^{\frac{{{\mu _E} + 1}}{2}}}}}{{{\kappa _E}^{\frac{{{\mu _E} - 1}}{2}}{{\overline {{\gamma _E}} }^{\frac{{{\mu _E} + 1}}{2}}}{e^{{\mu _E}{\kappa _E}}}}}\int\limits_0^\infty  {} {\gamma _E}^{\frac{{{\mu _E} - 1}}{2}}{e^{\frac{{ - {\mu _E}\left( {1 + {\kappa _E}} \right){\gamma _E}}}{{\overline {{\gamma _E}} }}}}{I_{{\mu _E} - 1}}\left( {2{\mu _E}\sqrt {\frac{{{\kappa _E}\left( {1 + {\kappa _E}} \right){\gamma _E}}}{{\overline {{\gamma _E}} }}} } \right)\\
 \times {Q_{{\mu _M}}}\left( {\sqrt {2{\kappa _M}{\mu _M}} ,\sqrt {\frac{{2\left( {1 + {\kappa _M}} \right){\mu _M}{\gamma _E}}}{{\overline {{\gamma _M}} }}} } \right)d{\gamma _E}.
 \label{eq:13}\end{multline}
\begin{IEEEproof}
 See Appendix \ref{app:C}.
\end{IEEEproof}

As can be seen, the solution obtained for $\eqref{eq:13}$ depends on the parameters $\mu_M$ and $\mu_E$ and thus the following propositions are valid.
\begin{prop} For the arbitrary real and positive  $\mu_M$ and $\mu_E$, $\eqref{eq:13}$ can be evaluated as, 
\begin{align}
{P_\alpha } &= \frac{{{{{{\beta _E}}}^{\frac{{{\mu _E} + 1}}{2}}}}}{{{\alpha _E}^{\frac{{{\mu _E} - 1}}{2}}{e^{{\alpha _E}}}}}\sum\limits_{k = 0}^\infty  {\sum\limits_{l = 0}^\infty  {\frac{{{{\left( {\sqrt {{\alpha _E}{\beta _E}} } \right)}^{{\mu _E} - 1 + 2k}}{{\left( {2{\alpha _M}} \right)}^l}{\beta _M}^{{\mu _M} + l}{\rm{\Gamma }}\left( {{\mu _E} + k + {\mu _M} + l} \right)}}{{k!{\rm{\Gamma }}\left( {{\mu _E} + k} \right)l!{\rm{\Gamma }}\left( {{\mu _M} + l} \right){2^{l}}{e^{{\alpha _M}}}\left( {{\mu _E} + k} \right){{\left( {{\beta _M} + {\beta _E}} \right)}^{{\mu _E} + k + {\mu _M} + l}}}}} } \nonumber \\
 & \times {_2}{F_1}\left( {1,{\mu _E} + k + {\mu _M} + l;{\mu _E} + k + 1;\frac{{{\beta _E}}}{{{\beta _M} + {\beta _E}}}} \right)
 \label{eq:14}\end{align}
where $a=  1/\bar\gamma_M$,  $b=  1/\bar\gamma_E$, ${\beta _M} = ({\kappa _M} + 1)a{\mu _M}$, ${\beta _E} = ({\kappa _E} + 1)b{\mu _E}$, ${\alpha _M} = {\kappa _M}{\mu _M}$, ${\alpha _E} = {\kappa _E}{\mu _E}$ and $\mathstrut_2 F_1(\cdot\ ,\cdot \ ;\ \cdot  \ ;\ \cdot)$  is the Gauss hypergeometric function~\cite{TofI}. \label{prop:1}
\end{prop}
\begin{IEEEproof}
 See Appendix \ref{app:D}.
\end{IEEEproof}

\begin{prop}
 For integer values of $\mu _M$ and $\mu _E$, an exact closed form solution for $\eqref{eq:13}$  is obtained as
\begin{multline}
{P_0} = 1 - \ $\'{P}$ \ + \ exp\left( { - \frac{{{A^2}r + {B^2}{r^{ - 1}}}}{{2R}}} \right)\sum\limits_{m =  - \mu }^v {{{\left( {\frac{A}{{Br}}} \right)}^m}{I_m}\left( {\frac{{AB}}{R}} \right)} \\
 \times \left\{ {\mathop \sum \limits_{k = 1}^\mu  \left( {\begin{array}{*{20}{c}}
{v + k}\\
{k + m}
\end{array}} \right){r^{v - k + 1}}{R^{ - v - k - 1}} - \mathop \sum \limits_{j = 1}^v \left( {\begin{array}{*{20}{c}}
j\\
m
\end{array}} \right){r^{j - 1}}{R^{ - j - 1}}} \right\}
 \label{eq:15}\end{multline}
where, 
\begin{align*}
\acute{P} &=  \ {Q_1}\left( {\sqrt {\frac{{2{\kappa _E}a{\mu _E}{\mu _M}\left( {1 + {\kappa _M}} \right)}}{{\left( {b\left( {1 + {\kappa _E}} \right){\mu _E}} \right) + \left( {a\left( {1 + {\kappa _M}} \right){\mu _M}} \right)}}} ,\sqrt {\frac{{2{\kappa _M}b{\mu _E}{\mu _M}\left( {1 + {\kappa _E}} \right)}}{{\left( {b\left( {1 + {\kappa _E}} \right){\mu _E}} \right) + \left( {a\left( {1 + {\kappa _M}} \right){\mu _M}} \right)}}} } \right)\\
 &- \left( {\frac{{b\left( {1 + {\kappa _E}} \right){\mu _E}}}{{\left( {b\left( {1 + {\kappa _E}} \right){\mu _E}} \right) + \left( {a\left( {1 + {\kappa _M}} \right){\mu _M}} \right)}}} \right)  exp\left( { - \frac{{a{\kappa _E}{\mu _M}{\mu _E}\left( {1 + {\kappa _M}} \right) + b{\kappa _M}{\mu _M}{\mu _E}\left( {1 + {\kappa _E}} \right)}}{{\left( {b\left( {1 + {\kappa _E}} \right){\mu _E}} \right) + \left( {a\left( {1 + {\kappa _M}} \right){\mu _M}} \right)}}} \right)\\
 & \times {I_0}\left( {\frac{{2{\mu _E}{\mu _M}\sqrt {ab{\kappa _M}{\kappa _E}\left( {1 + {\kappa _E}} \right)\left( {1 + {\kappa _M}} \right)} }}{{\left( {b\left( {1 + {\kappa _E}} \right){\mu _E}} \right) + \left( {a\left( {1 + {\kappa _M}} \right){\mu _M}} \right)}}} \right);
\end{align*}
$r = \sqrt {\frac{{\left( {1 + {\kappa _M}} \right){\mu _M}a}}{{\left( {1 + {\kappa _E}} \right){\mu _E}b}}}; $ $A = \sqrt {2{\kappa _E}{\mu _E}}; $ $B = \sqrt {2{\kappa _M}{\mu _M}}; $ $\mu = \mu_E -1 ;$ $v = \mu_M -1 ;$ and $R = r + r^{-1}$
\label{prop:2}\end{prop}
\begin{IEEEproof}
 See Appendix \ref{app:E}.
\end{IEEEproof}

%%%%%%%%%%%%%%%%%%%%%%%%%%%%%%%%%%%%%%%%%%%%%%%%%%%%%%%%%%%
\section{Special Cases and Numerical Results}
%%%%%%%%%%%%%%%%%%%%%%%%%%%%%%%%%%%%%%%%%%%%%%%%%%%%%%%%%%%

In this section, we verify the novel analytical and closed-form expressions of SPSC derived above by reducing them to a number of known special cases. We also use the formulations to provide a useful insight into the behavior of the SOP$^L$ and SPSC as a function of the fading parameters of the legitimate and eavesdroppers channels.

\subsection{Some Special Cases}
As discussed previously, because of the generality of the $\kappa$-$\mu$ fading model the results presented here encompass the probability of SPSC for a wide range of fading channels. For the reader's convenience a detailed list, along with the corresponding parameter values, is presented in Table \ref{Table:1}. We now discuss a few examples from Table \ref{Table:1}. 

\begin{table}[!t] 
\renewcommand{\arraystretch}{1.34}
\centering
\caption{Parameter Substitution to Obtain Secrecy Capacity of Special Cases Derived from the $\kappa$-$\mu$ Fading Model}
\label{Table:1}
\begin{tabular}{|p{7.4cm}|p{8cm}|}
\hline
    Scenario                                & Parameters to be substituted in $\eqref{eq:14}$ and/or $\eqref{eq:15}$                            \\ \hline \hline
  Rice / Rice                             & $\mu_M$ = $\mu_E$ = 1 ; $\kappa_M$ \textgreater \ 0 and $\kappa_E$ \textgreater \  0                      \\ \hline
  Nakagami-$m$ / Nakagami-$m$             & $\mu_M$ = $\mu_E$ \textgreater \ 0 ; $\kappa_M$ $\rightarrow$ \  0 and $\kappa_E$ $\rightarrow$ \  0        \\ \hline 
  Rayleigh / Rayleigh                     & $\mu_M$ = $\mu_E$ = 1 ; $\kappa_M$ $\rightarrow$ 0 and $\kappa_E$ $\rightarrow$ 0                           \\ \hline
  One-Sided Gaussian / One-Sided Gaussian & $\mu_M$ = $\mu_E$ = 0.5 ; $\kappa_M$ $\rightarrow$ 0 and $\kappa_E$ $\rightarrow$ 0                         \\ \hline
 $\kappa$-$\mu$ / Rice                   & $\mu_M$ \textgreater \ 0 ; $\kappa_M$ \textgreater \ 0 ; $\mu_E$ = 1 and  $\kappa_E$ \textgreater \ 0        \\ \hline
 $\kappa$-$\mu$ / Nakagami-$m$           & $\mu_M$ \textgreater \ 0 ; $\kappa_M$ \textgreater \ 0 ; $\mu_E$ \textgreater \ 0 and $\kappa_E$ $\rightarrow$ 0      \\ \hline
 $\kappa$-$\mu$ / Rayleigh               & $\mu_M$ \textgreater \ 0 ; $\kappa_M$ \textgreater \ 0 ; $\mu_E$ = 1 and $\kappa_E$ $\rightarrow$ 0          \\ \hline
 $\kappa$-$\mu$ / One-Sided Gaussian     & $\mu_M$ \textgreater \ 0 ; $\kappa_M$ \textgreater \ 0 ; $\mu_E$ = 0.5 and $\kappa_E$ $\rightarrow$ 0        \\ \hline
 Rice / $\kappa$-$\mu$                   & $\mu_M$ = 1 ; $\kappa_M$ \textgreater \ 0 ; $\mu_E$ \textgreater \ 0 and $\kappa_E$ \textgreater \ 0               \\ \hline
 Rice / Rayleigh                         & $\mu_M$ = 1 ; $\kappa_M$ \textgreater \ 0 ; $\mu_E$ = 1 and $\kappa_E$ $\rightarrow$ 0                         \\ \hline
 Rice / Nakagami-$m$                    & $\mu_M$ = 1 ; $\kappa_M$ \textgreater \ 0 ; $\mu_E$ \textgreater \ 0 and $\kappa_E$ $\rightarrow$ 0              \\ \hline
 Rice/ One-Sided Gaussian                & $\mu_M$ = 1 ; $\kappa_M$ \textgreater \ 0 ; $\mu_E$ = 0.5 and $\kappa_E$ $\rightarrow$ 0                       \\ \hline
 Nakagami-$m$ / $\kappa$-$\mu$           & $\mu_M$ \textgreater \ 0 and $\kappa_M$ $\rightarrow$ 0 ; $\mu_E$ \textgreater \ 0 and $\kappa_E$ \textgreater \ 0 \\ \hline
 Nakagami-$m$ / Rice                     & $\mu_M$ \textgreater \ 0 and $\kappa_M$ $\rightarrow$ 0 ; $\mu_E$ = 1 and $\kappa_E$ \textgreater \ 0            \\ \hline
 Nakagami-$m$ / Rayleigh                 & $\mu_M$ \textgreater \ 0 and $\kappa_M$ $\rightarrow$ 0 ; $\mu_E$ = 1 and $\kappa_E$ $\rightarrow$ 0           \\ \hline
 Nakagami-$m$ / One-Sided Gaussian       & $\mu_M$ \textgreater \ 0 and $\kappa_M$ $\rightarrow$ 0 ; $\mu_E$ = 0.5 and $\kappa_E$ $\rightarrow$ 0         \\ \hline
 Rayleigh / $\kappa$-$\mu$               & $\mu_M$ = 1 ; $\kappa_M$ $\rightarrow$ 0 ; $\mu_E$ \textgreater \ 0 and $\kappa_E$ \textgreater~0              \\ \hline
 Rayleigh / Rice                         & $\mu_M$ = 1 ; $\kappa_M$ $\rightarrow$ 0 ; $\mu_E$ = 1 and $\kappa_E$ \textgreater \ 0                         \\ \hline
 Rayleigh / Nakagami-$m$                 & $\mu_M$ = 1 ; $\kappa_M$ $\rightarrow$ 0 ; $\mu_E$ \textgreater \ 0 and $\kappa_E$ $\rightarrow$ 0             \\ \hline
 Rayleigh / One-Sided Gaussian           & $\mu_M$ = 1 ; $\kappa_M$ $\rightarrow$ 0 ; $\mu_E$ = 0.5 and $\kappa_E$ $\rightarrow$ 0                      \\ \hline
 One-Sided Gaussian /~$\kappa$-$\mu$      & $\mu_M$ = 0.5 ; $\kappa_M$ $\rightarrow$ 0 ; $\mu_E$ \textgreater \ 0 and $\kappa_E$ \textgreater~0            \\ \hline
 One-Sided Gaussian / Rice               & $\mu_M$ = 0.5 ; $\kappa_M$ $\rightarrow$ 0 ; $\mu_E$ = 1 and $\kappa_E$ \textgreater \ 0                       \\ \hline
 One-Sided Gaussian / Rayleigh           & $\mu_M$ = 0.5 ; $\kappa_M$ $\rightarrow$ 0 ; $\mu_E$ = 1 and $\kappa_E$ $\rightarrow$ 0                      \\ \hline
 One-Sided Gaussian / Nakagami-$m$       & $\mu_M$ = 0.5 ; $\kappa_M$ $\rightarrow$ 0 ; $\mu_E$ \textgreater \ 0 and $\kappa_E$ $\rightarrow$ 0           \\ \hline
\end{tabular}
\end{table}

%%%%%%%%%%%%%%%%%%%%%%%%%%%%%%%%%%%%%%%%%%%%%%%%%%%%%%%%%%%
\subsubsection{Rice / Rice and Rayleigh / Rayleigh}
\begin{figure}
	\centering
	 \includegraphics[width=0.6\textwidth]{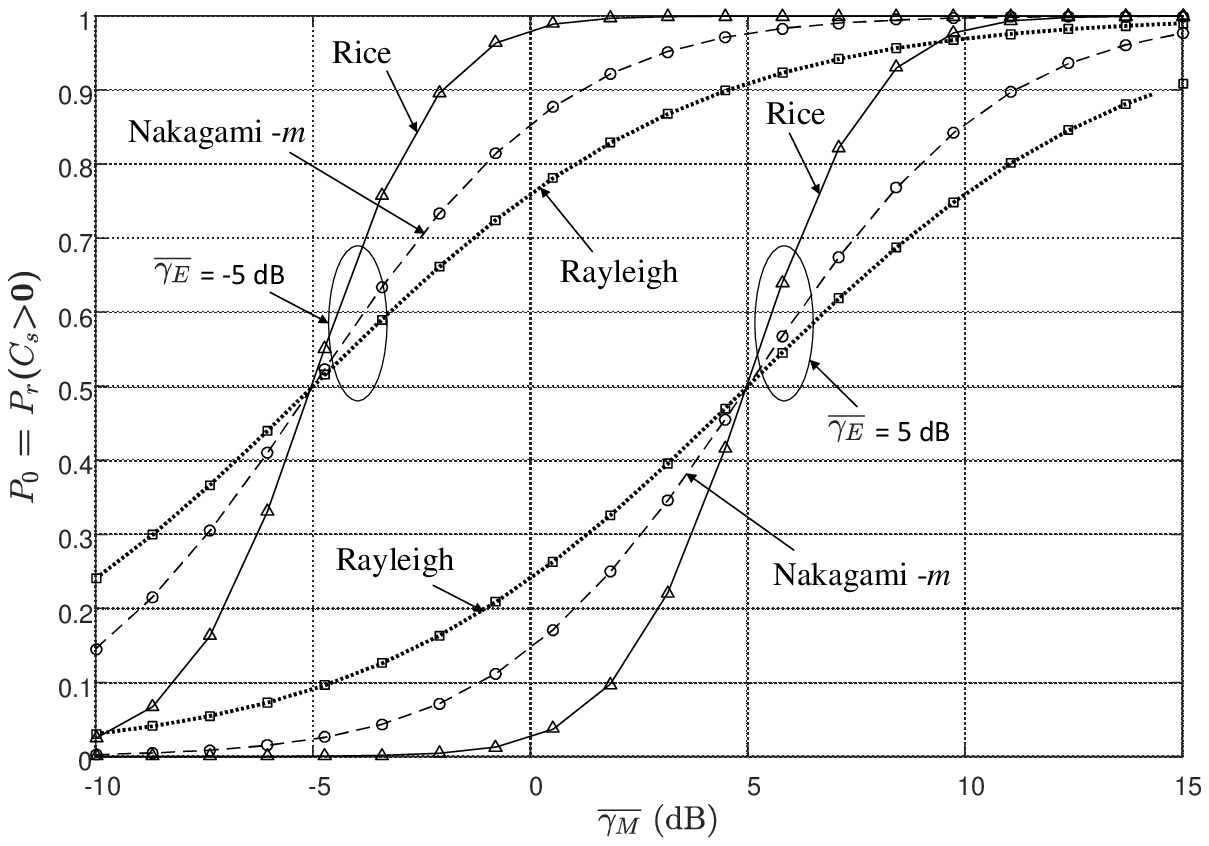}
	\caption{The probability of non-zero secrecy capacity versus $\bar\gamma_M$. Lines represent equation $\eqref{eq:14}$ and $\eqref{eq:15}$ for the special case of Rice ($\mu_M$ = $\mu_E$ = 1; $\kappa_M$ = 15, $\kappa_E$ = 12); Nakagami-$m$  ($\kappa_M$ $\rightarrow$ 0 and $\kappa_E$ $\rightarrow$ 0; $\mu_M$ = $\mu_E$ = 2) and Rayleigh fading ($\kappa_M$ $\rightarrow$ 0 and $\kappa_E$ $\rightarrow$ 0; $\mu_M$ = $\mu_E$ = 1). Triangle markers represent~\cite[eq. 10]{6338984} with $\kappa_a$ = 15, $\kappa_b$ = 12 for Rice;  circle markers~\cite[eq. 8]{nak5469979multi} with $m$ = 2; N = 1 for Nakagami-$m$  and square markers~\cite[eq. 5]{barros2006secrecy} for Rayleigh fading. }
	\label{fig:img-2}
\end{figure}

%%%%%%%%%%%%%%%%%%%%%%%%%%%%%%%%%%%%%%%%%%%%%%%%%%%%%%%%%%%
As shown in Table \ref{Table:1}, to obtain the probability of SPSC for the case when both the main channel and eavesdropper's channel undergo Rician fading (i.e. a Rice / Rice fading scenario), we substitute $\mu_M$ =  $\mu_E$ = 1 into $\eqref{eq:14}$ and/or $\eqref{eq:15}$ in which case these reduce to
\begin{multline}
{P_0} = \frac{{\left( {1 + {\kappa _E}} \right)}}{{\overline {{\gamma _E}} \ {e^{{\kappa _E}}}}}\sum\limits_{k = 0}^\infty  {\sum\limits_{l = 0}^\infty  {\frac{{{{\left( {\sqrt {{\kappa _E}\left( {1 + {\kappa _E}} \right)b} } \right)}^{2k}}}}{{k!{\rm{\Gamma }}\left( {1 + k} \right)}}\frac{{{{\left( {2{\kappa _M}} \right)}^l}}}{{l!{\rm{\Gamma }}\left( {1 + l} \right){2^{l}}{e^{{\kappa _M}}}}}} } \\
\times \frac{{\left( {({\kappa _M} + 1} \right)a{)^{1 + l}}{\rm{\Gamma }}\left( {2 + k + l} \right)}}{{\left( {1 + k} \right){{\left( {\left( {{\kappa _M} + 1} \right)a + \left( {{\kappa _E} + 1} \right)b} \right)}^{2 + k + l}}}} \ {_2}{F_1}\left( {1,2 + k + l;2 + k;\frac{{\left( {{\kappa _E} + 1} \right)b}}{{\left( {{\kappa _M} + 1} \right)a + \left( {{\kappa _E} + 1} \right)b}}} \right)
\label{eq:16}\end{multline}
and
\begin{multline}
{P_0} = 1 - {Q_1}\left( {\sqrt {\frac{{2{\kappa _E}a\left( {1 + {\kappa _M}} \right)}}{{b\left( {1 + {\kappa _E}} \right) + a\left( {1 + {\kappa _M}} \right)}}} ,\sqrt {\frac{{2{\kappa _M}b\left( {1 + {\kappa _E}} \right)}}{{b\left( {1 + {\kappa _E}} \right) + a\left( {1 + {\kappa _M}} \right)}}} } \right) + \\
\left( {\frac{{b\left( {1 + {\kappa _E}} \right)}}{{b\left( {1 + {\kappa _E}} \right) + a\left( {1 + {\kappa _M}} \right)}}} \right){e^{\left( { - \frac{{a{k_E}\left( {1 + {\kappa _M}} \right) + b{\kappa _M}\left( {1 + {\kappa _E}} \right)}}{{b\left( {1 + {\kappa _E}} \right) + a\left( {1 + {\kappa _M}} \right)}}} \right)}} \ {I_0}\left( {\frac{{2\sqrt {ab{\kappa _M}{\kappa _E}\left( {1 + {\kappa _E}} \right)\left( {1 + {\kappa _M}} \right)} }}{{b\left( {1 + {\kappa _E}} \right) + a\left( {1 + {\kappa _M}} \right)}}} \right)
\label{eq:17}\end{multline}
which are in exact agreement with the result reported in~\cite[eq. 10]{6338984} and are illustrated visually in \figref{fig:img-2}. Of course letting $\kappa_M$ = $\kappa_E$ = 0, then
\begin{equation}
{P_0} = \frac{{\overline {{\gamma _M}} }}{{\overline {{\gamma _M}}  + \overline {{\gamma _E}} }}
\label{eq:18}\end{equation}
which matches exactly with that given in~\cite[eq. 5]{barros2006secrecy} for a Rayleigh / Rayleigh fading scenario.

%%%%%%%%%%%%%%%%%%%%%%%%%%%%%%%%%%%%%%%%%%%%%%%%%%%%%%%%%%%
\subsubsection{Nakagami-$m$  / Nakagami-$m$}
%%%%%%%%%%%%%%%%%%%%%%%%%%%%%%%%%%%%%%%%%%%%%%%%%%%%%%%%%%%
By letting $\kappa_M$ $\rightarrow$ 0 and $\kappa_E$ $\rightarrow$ 0 into $\eqref{eq:14}$ and/or $\eqref{eq:15}$, we obtain the probability of SPSC for the scenario where both the legitimate and non-legitimate users channels undergo Nakagami-$m$ fading. As shown in \figref{fig:img-2}, our results have been compared with that reported in~\cite[eq. 8]{nak5469979multi}. It should be noted that the expression proposed in~\cite[eq. 8]{nak5469979multi} is valid only for identical fading parameters of the main and the eavesdroppers channels and for integer values of the shape parameter, $m$ (or equivalently  $\mu$ when $\kappa$ $\rightarrow$ 0) when a single eavesdropper is considered whereas the equation proposed here is valid for any positive real value of the $\mu$ parameter.  As shown in \figref{fig:img-2}, for the case of integer $\mu_M$ and $\mu_E$, the results presented here are in exact agreement with those presented in~\cite{nak5469979multi}.

%%%%%%%%%%%%%%%%%%%%%%%%%%%%%%%%%%%%%%%%%%%%%%%%%%%%%%%%%%%
\subsubsection{Other Fading Scenarios}
%%%%%%%%%%%%%%%%%%%%%%%%%%%%%%%%%%%%%%%%%%%%%%%%%%%%%%%%%%%
\begin{figure}[h]
	\centering
	\centering
		\includegraphics[width=0.6\textwidth]{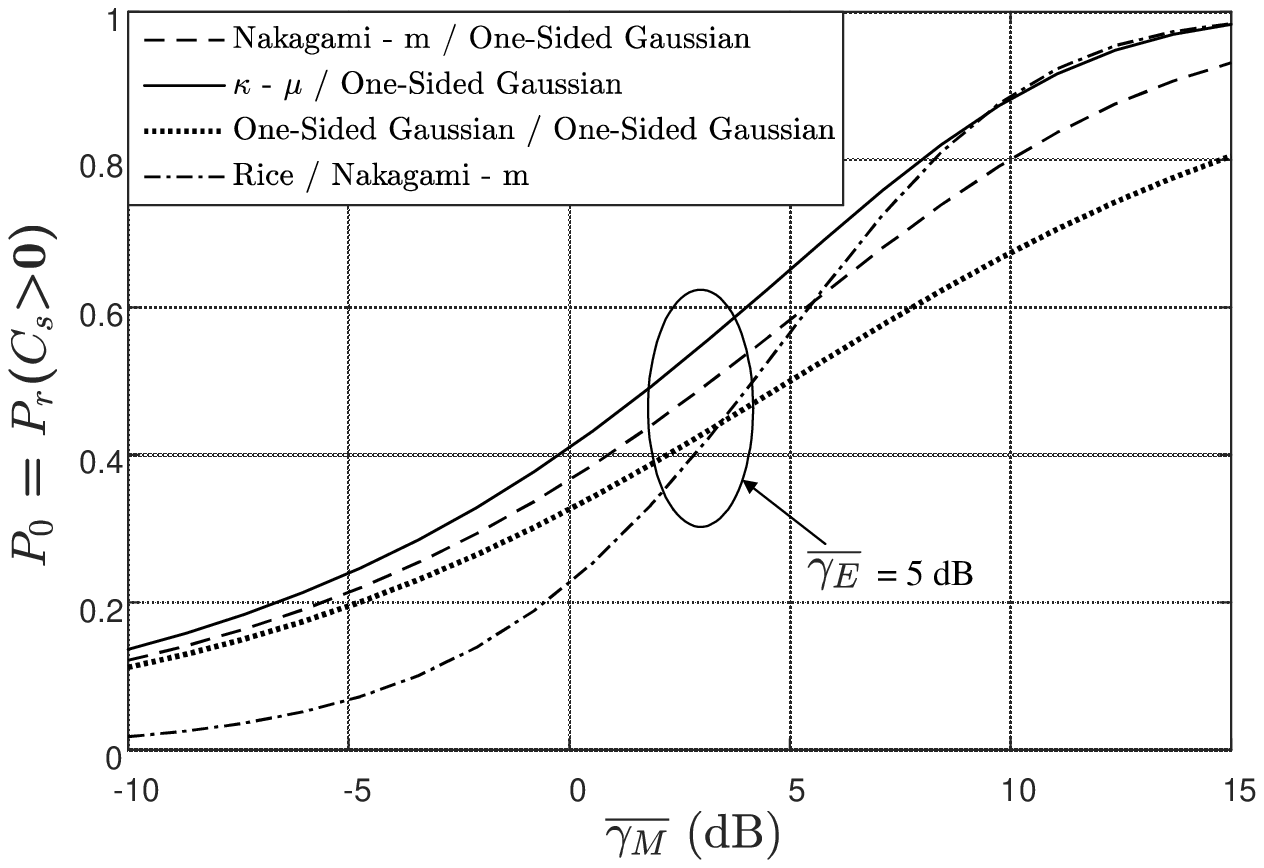}
	\caption{The probability of non-zero secrecy capacity versus $\bar\gamma_M$ for the special case of Nakagami-$m$ / One-Sided Gaussian ($\kappa_M$ $\rightarrow$ 0; $\mu_M$ = 1.2; $\mu_E$ = 0.5 and $\kappa_E$ $\rightarrow$ 0), $\kappa$-$\mu$ / One-Sided Gaussian ($\kappa_M$ = 5; $\mu_M$ = 1.2; $\kappa_E$ $\rightarrow$ 0 and $\mu_E$ = 0.5), One-Sided Gaussian / One-Sided Gaussian ($\mu_M$ = $\mu_E$ = 0.5; $\kappa_M$ $\rightarrow$ 0 and $\kappa_E$ $\rightarrow$ 0) and Rice / Nakagami-$m$ ($\kappa_M$ = 5; $\mu_M$ = 1; $\kappa_E$ $\rightarrow$ 0 and $\mu_E$ = 1.2)  }
	\label{fig:img-3}
\end{figure}
In a similar manner, the probability of non-zero secrecy capacity for several different fading combinations, most of which have not been reported previously in the open literature, can be obtained from Table \ref{Table:1}. \figref{fig:img-3} shows the shape of probability of SPSC for a selection of these scenarios, namely the Nakagami-$m$ / One-Sided Gaussian ($\kappa_M$ $\rightarrow$ 0 ; $\mu_M$ = 1.2 ;  $\mu_E$ = 0.5 and  $\kappa_E$ $\rightarrow$ 0), $\kappa$-$\mu$ / One-Sided Gaussian ($\mu_M$ = 1.2 ; $\kappa_M$ = 5 ;  $\mu_E$ = 0.5 and $\kappa_E$ $\rightarrow$ 0) , One-Sided Gaussian / One-Sided Gaussian ($\mu_M$ = $\mu_E$ = 0.5 ; $\kappa_M$ $\rightarrow$ 0 and $\kappa_E$ $\rightarrow$ 0) and Rice / Nakagami-$m$ ($\mu_M$ = 1 ; $\kappa_M$ = 5 ; $\mu_E$ = 1.2 and $\kappa_E$ $\rightarrow$ 0 ).

%%%%%%%%%%%%%%%%%%%%%%%%%%%%%%%%%%%%%%%%%%%%%%%%%%%%%%%%%%%
\subsection{Numerical Results}
%%%%%%%%%%%%%%%%%%%%%%%%%%%%%%%%%%%%%%%%%%%%%%%%%%%%%%%%%%%

%\begin{figure}[h]
%\centering
%\includegraphics[trim=0 6cm 0 8cm, clip=true, totalheight=0.4\textheight, angle=0]{img_41.eps}
%\caption{The probability of non-zero secrecy capacity with $\kappa_M = 4$ }
%\label{fig:img-41}
%\end{figure}

\begin{figure}[h]
	\centering
		\includegraphics[width=1\textwidth]{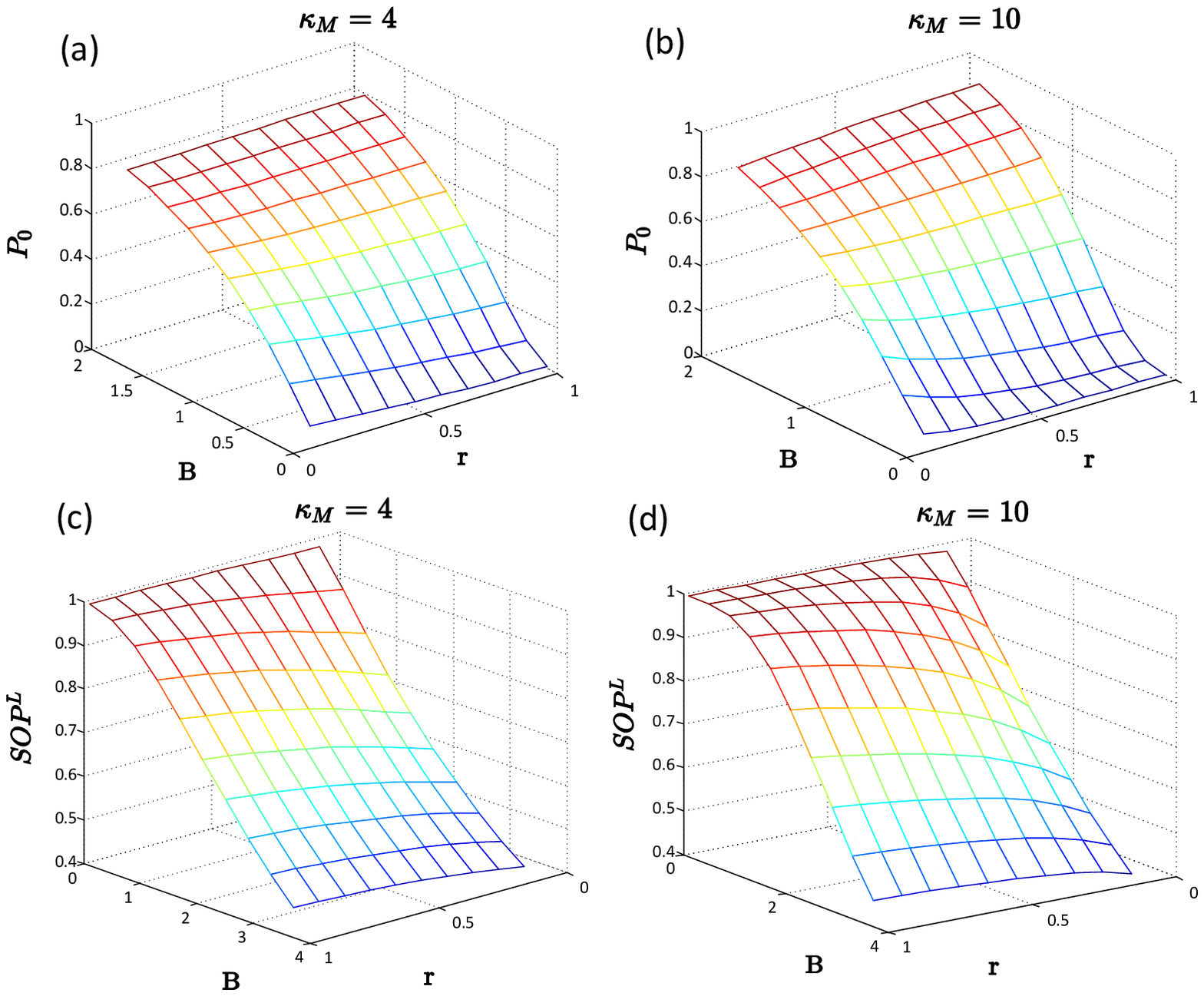}
	\caption{The probability of non-zero secrecy capacity with (a) $\kappa_M = 4$ and (b) $\kappa_M = 10$, and the secrecy outage probability with (c) $\kappa_M = 4$ and (d) $\kappa_M = 10$.}
	\label{fig:img-41}
\end{figure}

Here, we discuss the behavior of $P_0$ and SOP$^L$ as a function of parameters \{$\kappa_M$, $\mu_M$, $\bar\gamma_M$\} and \{$\kappa_E$, $\mu_E$, $\bar\gamma_E$\}. Let $B$ represent the ratio of the average SNR of the main channel to that of the eavesdropper's channel i.e, $B = \bar\gamma_M/\bar\gamma_E$ and $r$ represent the ratio, $\kappa_E/\kappa_M$. For the figures shown in this section, we set $R_S = 1$~dB, $\mu_M = 1.4$ and $\mu_E = 1.2$. Four  example  profiles for the probability of SPSC and SOP$^L$ are  illustrated in Fig. \ref{fig:img-41}.

\vspace{10 mm}

%\begin{figure}[h]
%\centering
%\includegraphics[trim=0 6cm 0 8cm, clip=true, totalheight=0.4\textheight, angle=0]{img_42.eps}
%\caption{The probability of non-zero secrecy capacity with $\kappa_M = 10$ }
%\label{fig:img_42}
%\end{figure}
%
%\begin{figure}[h]
%\centering
%\includegraphics[trim=0 6cm 0 8cm, clip=true, totalheight=0.4\textheight, angle=0]{img_43.eps}
%\caption{The secrecy outage probability with $\kappa_M = 4$ }
%\label{fig:img_43}
%\end{figure}

From these figures, we observe that the probability of SPSC increases and the secrecy outage probability decreases as $B$ increases. This is because larger values of $B$ indicate that the signal quality of the main channel is better than that of the eavesdropper's channel. From Figs. \ref{fig:img-41}(a) and (b), we observe that the probability of SPSC is non-zero even when $B < 1$. Furthermore, $P_0$ increases as $\kappa_M$, the main channel's fading parameter, increases, while for a fixed $\kappa_M$ with $\bar\gamma_M$ \textless \ $\bar\gamma_E$, $P_0$ increases as $\kappa_E$ decreases. It can be seen from Figs. \ref{fig:img-41}(c) and (d) that the SOP decreases as $\kappa_M$ increases. Additionally, for a fixed $\kappa_M$ and when $B < 1$, the SOP decreases as $\kappa_E$ decreases. Similar observations are made when $\kappa_M$ and $\kappa_E$ are set and $\mu_M$ and $\mu_E$ are varied.

%\begin{figure}[h]
%\centering
%\includegraphics[trim=0 6cm 0 8cm, clip=true, totalheight=0.4\textheight, angle=0]{img_44.eps}
%\caption{The secrecy outage probability with $\kappa_M = 10$ }
%\label{fig:img_44}
%\end{figure}

%%%%%%%%%%%%%%%%%%%%%%%%%%%%%%%%%%%%%%%%%%%%%%%%%%%%%%%%%%%
\section{Applications of SPSC for Devices Operating in $\kappa$-$\mu$ Fading Channels}
%%%%%%%%%%%%%%%%%%%%%%%%%%%%%%%%%%%%%%%%%%%%%%%%%%%%%%%%%%%
To illustrate the utility of the new equations proposed here, we now analyze the probability of SPSC for a number of emerging applications such as cellular device-to-device, body area network (BAN) and vehicle-to-vehicle (V2V) communications using channel data obtained from field trials. For all of the measurements conducted in this study, we considered a three node system which consisted of Alice, herein denoted node A, which acted as the transmitter and also Bob (node B) and Eve (node E) which acted as the receivers. Each of the nodes, A, B and E, consisted of an ML5805 transceiver, manufactured by RFMD. The transceiver boards were interfaced with a PIC32MX which acted as a baseband controller and allowed the analog received signal strength (RSS) to be sampled with a 10-bit quantization depth.  For all of the experiments conducted here, node A was configured to output a continuous wave signal with a power level of +21 dBm at 5.8 GHz while nodes B and E sampled the channel at a rate of 1 kHz. The antennas used by the transmitter and the receivers were +2.3 dBi sleeve dipole antennas (Mobile Mark model PSKN3-24/55S).

%%%%%%%%%%%%%%%%%%%%%%%%%%%%%%%%%%%%%%%%%%%%%%%%%%%%%%%%%%%
\subsection{Device-to-Device Scenario}
%%%%%%%%%%%%%%%%%%%%%%%%%%%%%%%%%%%%%%%%%%%%%%%%%%%%%%%%%%%
The first set of measurements considered cellular device-to-device communications channels operating at 5.8 GHz in an indoor environment. The experiments were conducted in a large seminar room located on the first floor of the ECIT building at Queen's University Belfast in the United Kingdom. The seminar room had dimensions of 7.92 m x 12.58 m x 2.75 m and contained a number of chairs, some desks constructed from medium density fiberboard, a projector and a white board. For these measurements, the antennas were housed in a compact acrylonitrile butadiene styrene (ABS) enclosure (107 x 55 x 20 mm). This setup was representative of the form factor of a smart phone which allowed the user to hold the device as they normally would to make a voice call. Each antenna was securely fixed to the inside of the enclosure using a small strip of Velcro\textsuperscript{\textregistered}. The antennas were connected using low-loss coaxial cables to nodes A, B and E. 

\begin{figure}
	\centering
	 \includegraphics[width=0.55\textwidth]{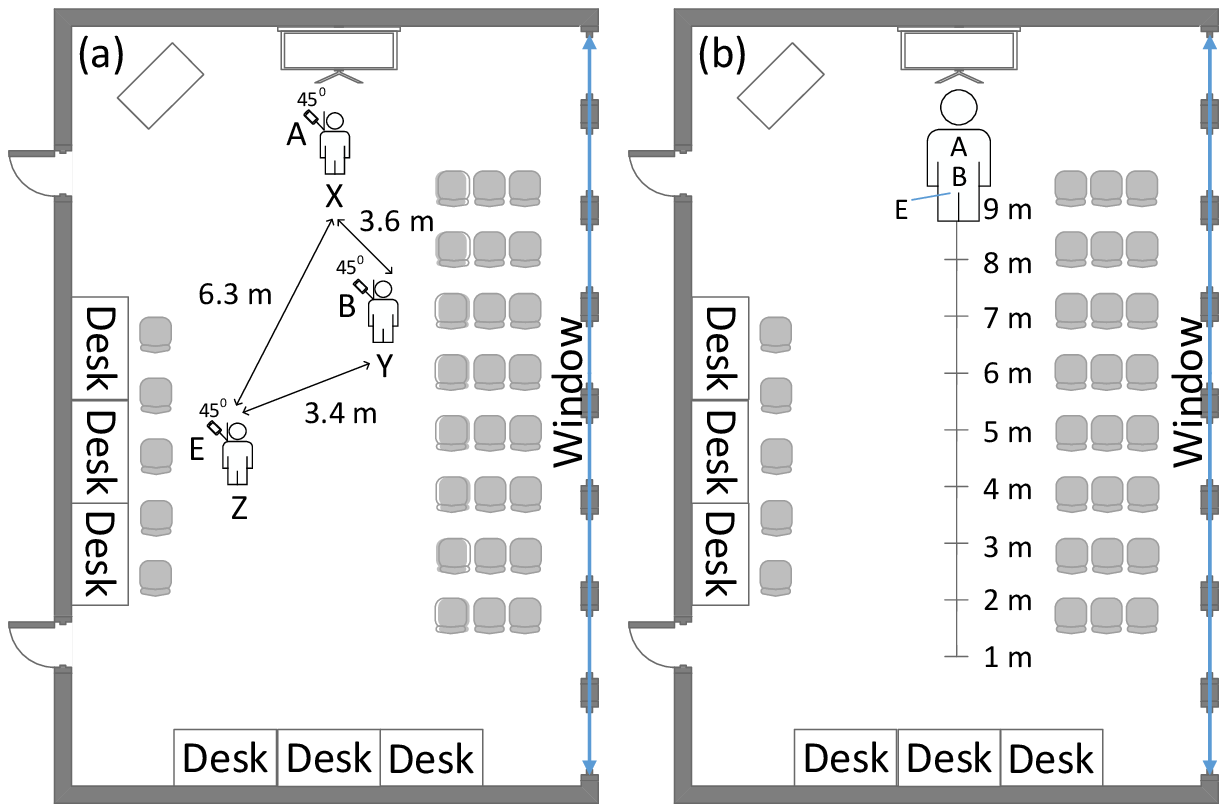}
	\caption{Seminar room environment showing the position of nodes A, B and E for (a) D2D scenario (b) on-body scenario.}
	\label{fig:img-7}
\end{figure}

The experiment was performed when the room was unoccupied except for the test subjects holding the devices. As shown in Fig. \ref{fig:img-7}(a), this particular scenario considered three persons carrying nodes A, B and E who were positioned at points X, Y and Z respectively. The three test subjects using nodes A, B and E were adult males of height 1.72 m, 1.84 m and 1.83 m; mass 80 kg, 92 kg and 74 kg, respectively. During the measurement trial, all three persons were initially stationary and had the hypothetical user equipment (UE) positioned at their heads. The persons at points Y and Z were then instructed to move around randomly within a circle of radius 0.5 m from their starting points while imitating a voice call. A total of 74763 samples of the received signal power were obtained and used for parameter estimation.

Figs. \ref{fig:img-9}(a) and (b) show the empirical PDF of the signal envelope for Bob and Eve compared to the $\kappa$-$\mu$ PDF given in ~\cite[eq. 11]{4231253} for the D2D channel measurements. All parameter estimates for the $\kappa$-$\mu$ fading model were obtained using the \texttt{lsqnonlin} function available in the Optimization toolbox of Matlab along with the $\kappa$-$\mu$ PDF given in ~\cite[eq. 11]{4231253}. It should be noted that to remove the impact of any shadowing processes, which are not accounted for in the $\kappa$-$\mu$ fading model, the data sets were normalized to their respective local means prior to parameter estimation. To determine the window size for extraction of the local mean signal, the raw data was visually inspected and overlaid with the local mean signal for differing window sizes. For the D2D channel data, a smoothing window of 500 samples was considered. As we can quite clearly see, from Figs. \ref{fig:img-9}(a) and (b), the envelope PDF of the $\kappa$-$\mu$ fading model provides an excellent fit to the D2D data. To allow the reader to reproduce these plots, parameter estimates for all three measurement scenarios are given in Table \ref{Table:2}. 

Using the parameter estimates obtained from the field trials, Figs. \ref{fig:img-9}(c) depicts the estimated probability of SPSC versus $\bar\gamma_M$ for selected values of $\bar\gamma_E$ for the measured D2D channel.  In this instance, it can be seen that the estimates for $\kappa$ of the main and eavesdropper's channels are comparable and also greater than 0, suggesting that a dominant component existed for both. We also observe that the parameter estimates for $\mu_M$ and $\mu_E$ are both quite close to 1, suggesting that a single multipath cluster contributes to the signal received by both Bob (node B) and Eve (node E) and thus this fading scenario is quite close to the Rician case. For this fading environment, we note that $P_0$ increases as the average SNR of the main channel increases. Furthermore, for a fixed $\bar\gamma_M$, the channel becomes more susceptible to eavesdropping as the average SNR of the eavesdropper's channel increases.
%%\begin{figure*}[!t]
%     \centering {$\kappa_M$, $\mu_M$, $\bar\gamma_M$\} and \{$\kappa_E$, $\mu_E$, $\bar\gamma_E$\}
%     \includegraphics[width=0.8\linewidth, height =0.4\linewidth]{kappa_mu_nak_mu_m_mu_e_m_4.eps}
%     \caption{Test Image}
%     \label{fig1_test}
%%\end{figure*}%

%%%%%%%%%%%%%%%%%%%%%%%%%%%%%%%%%%%%%%%%%%%%%%%%%%%%%%%%%%%
\subsection{Body Area Network Scenario}
%%%%%%%%%%%%%%%%%%%%%%%%%%%%%%%%%%%%%%%%%%%%%%%%%%%%%%%%%%%
The second set of measurements considered on-body communications channels operating at 5.8 GHz as found in body area networks. The experiment was performed in the same seminar room discussed above which was unoccupied except for the test subject on which the on-body nodes were placed. For this scenario, to maximize coupling across the body surface, the antennas were mounted normal to the torso of an adult male of height 1.83 m and mass 74 kg. Node A was positioned at the front central chest region at a height of 1.42 m while nodes B and E were placed on the rear of the test subject at the central waist region at a height of 1.15 m and the right back-pocket at a height of 0.92 m, respectively.  The measurements considered the case when the hypothetical BAN user walked along a straight line within the large room, covering a total distance of 9 m as shown in Fig. \ref{fig:img-7}(b). For the BAN scenario, a total of 19260 samples of the received signal power were obtained and used for parameter estimation.

Figs. \ref{fig:img-9}(d) and (e) show the empirical PDF of the signal envelope for Bob and Eve, again compared to the $\kappa$-$\mu$ PDF given in ~\cite[eq. 11]{4231253}. Identical to the analysis of the D2D measurements, the optimum window size was determined from the raw channel data. In this case a smoothing window of 100 samples was used. Again, the $\kappa$-$\mu$ PDF was found to provide an excellent fit to the empirical data for both Bob and Eve. Interestingly for the BAN configuration considered here, the estimated $\kappa$ parameter of the eavesdropper's channel was greater than that of the main channel, while for the estimated $\mu$ parameters, the converse situation was true (Table \ref{Table:2}). Fig. \ref{fig:img-9}(f) shows the probability of SPSC versus $\bar\gamma_M$ with selected values of $\bar\gamma_E$ for the measured BAN channel. It can be quite clearly seen that for fixed $\bar\gamma_E$, the probability of SPSC improves as the legitimate channel experiences a higher average SNR.

%%%%%%%%%%%%%%%%%%%%%%%%%%%%%%%%%%%%%%%%%%%%%%%%%%%%%%%%%%%
\subsection{Vehicle-to-Vehicle Scenario}
\begin{figure}[htbp]
	\centering
		\includegraphics{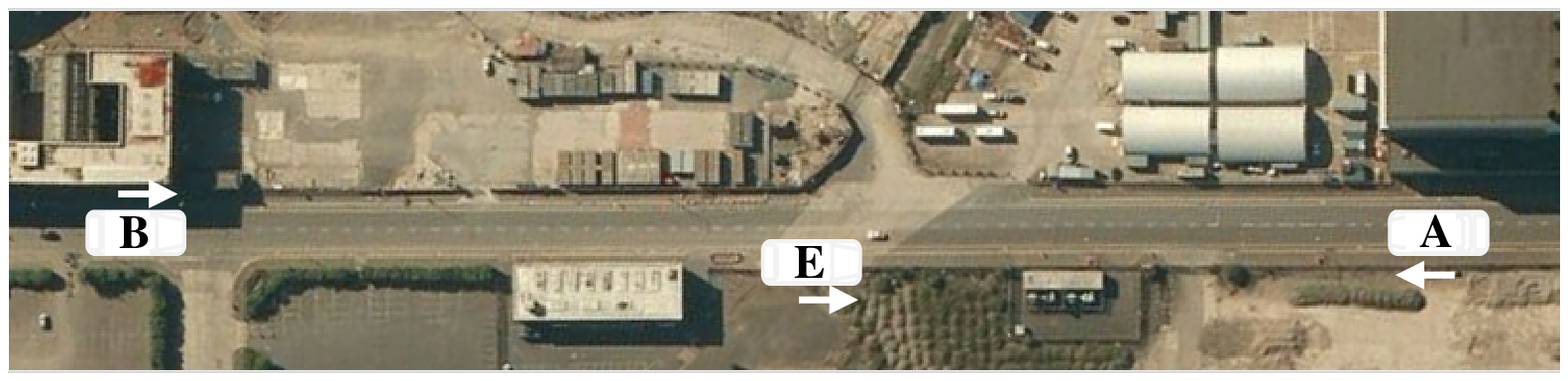}
	\caption{Satellite view of measurement environment showing the position of nodes A, B and E for V2V scenario. The vehicle with node E remained parked on the side of the road and was oriented such that it faced directly towards node A. Arrows indicate the front of the car. }
	\label{fig:img-8}
\end{figure}

%%%%%%%%%%%%%%%%%%%%%%%%%%%%%%%%%%%%%%%%%%%%%%%%%%%%%%%%%%%
The third set of measurements considered vehicle-to-vehicle communication channels operating at 5.8 GHz. The experiments were conducted in a business district environment in the Titanic Quarter of Belfast, United Kingdom. As shown in Fig. \ref{fig:img-8}, the area consisted of a straight road with  a number of office buildings nearby. For this particular scenario, nodes A, B and E were placed on the center of the dash boards of three different vehicles; namely, a Vauxhall (Opel in Europe) Zafira SRi, a  Vauxhall Astra SRi and a Hyundai Getz. The initial positions of the vehicles are shown in Fig. \ref{fig:img-8}. The measurements began when the vehicles that contained nodes A and B started approaching one another at a speed of 30mph. During these measurements the vehicle containing node E remained parked (with the driver still seated inside) on the side of the road as indicated in Fig. \ref{fig:img-8}. It should be noted that all of the channel measurements made in this scenario were performed during off-peak traffic hours and were subject to perturbations caused by the driver, movement of the nearby pedestrians and other vehicular traffic. A total of 56579 samples of the received signal power were obtained and used for parameter estimation.

\begin{table}[!t] 
\renewcommand{\arraystretch}{1.1}
\centering
\caption{Parameter Estimates for the $\kappa-\mu$ Fading Model Obtained from the Field Measurements}
\label{Table:2}
\begin{tabular}{|p{3cm}|p{1cm}|p{1cm}|p{1cm}|p{1cm}|p{1cm}|p{1cm}|}
\hline
    Fading Channel                        & $\hat{\kappa}_M$   & $\hat{\mu}_M$   & $\hat{\bar{r}}_M$   & $\hat{\kappa}_E$   & $\hat{\mu}_E$   & $\hat{\bar{r}}_E$           
		\\ \hline
  D2D                                     & 1.07   & 0.91   & 1.22   & 1.11   & 0.92   & 1.19                 \\ \hline
 On-Body                                  & 2.92	 & 0.75	  & 1.17	 & 3.60	  & 0.67	 & 1.17                  \\ \hline 
  V2V                                     &5.02	   &0.70	  & 1.04	 & 7.17	  & 0.60	 & 1.03                   \\ \hline     
\end{tabular}
\end{table}

\begin{figure}[htbp]
	\centering
		\includegraphics{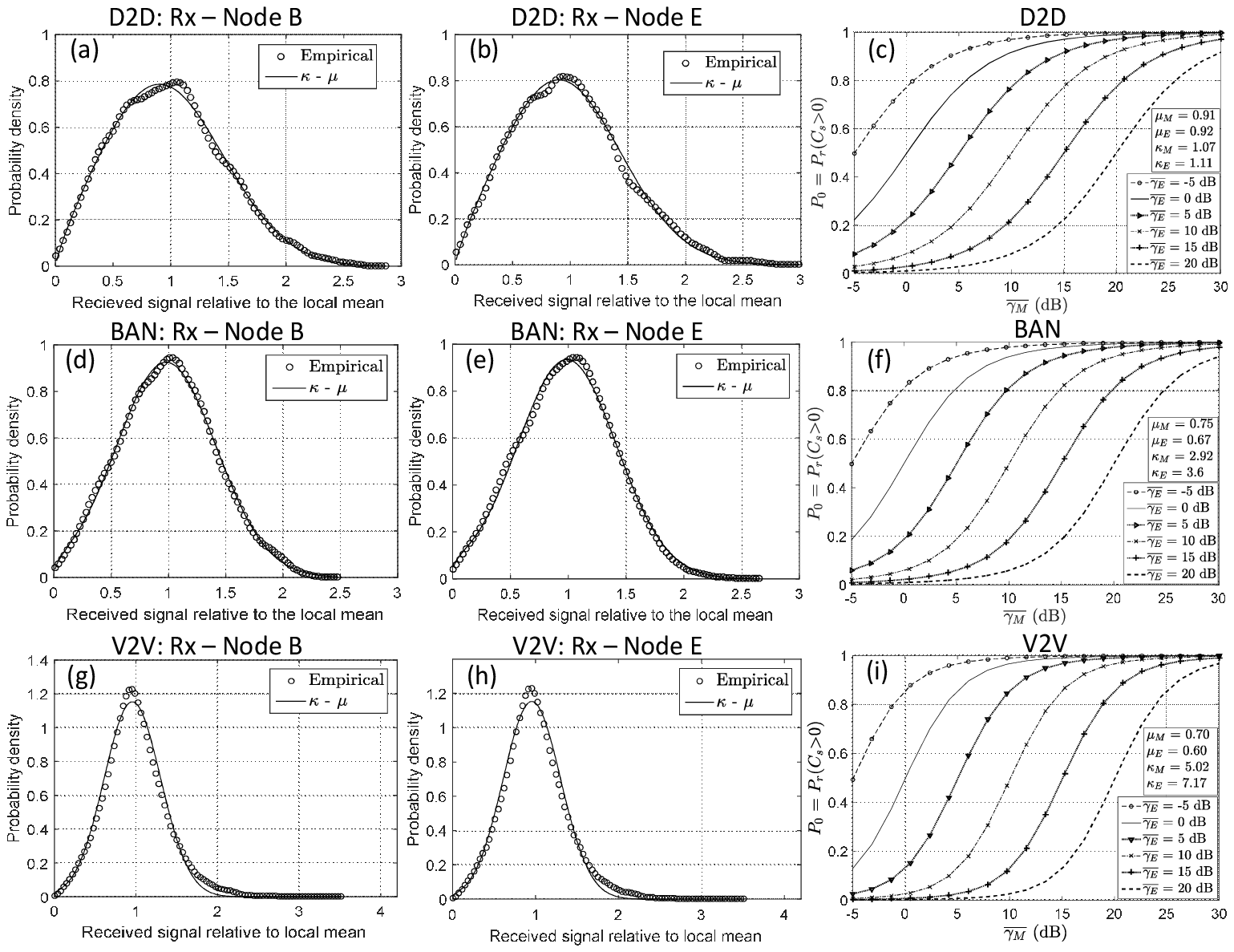}
	\caption{Empirical envelope PDF of node B and E compared to the $\kappa$-$\mu$ PDF given in~\cite[eq. 11]{4231253} and the probability of SPSC versus $\bar\gamma_M$ with selected values of $\bar\gamma_E$ for D2D, BAN and V2V channel measurements, respectively.  }
	\label{fig:img-9}
\end{figure}

Figs. \ref{fig:img-9}(g) and (h) show the empirical PDF of the signal envelope for Bob and Eve compared to the $\kappa$-$\mu$ PDF. Similar to the analysis of the D2D and BAN measurements, the optimum window size was determined from the raw channel data. The local mean for the V2V measurements was calculated over 200 samples. From these figures we can see that the PDF of the $\kappa$-$\mu$ fading model provides a very good approximation to the V2V data. From Table \ref{Table:2}, it can be seen that the estimates of $\kappa$ for the main and the eavesdropper's channels are greater than 0 whilst the estimated $\mu$ parameters are less than 1, suggesting that a dominant component exists and that these channels suffer less from multipath. We also observe that the estimated $\kappa$ parameter of the eavesdropper's channel was greater than that of the legitimate channel, the estimated $\mu$ parameter of the main channel was only marginally greater than that of the eavesdropper's channel. Fig. \ref{fig:img-9}(i) depicts the probability of SPSC versus $\bar\gamma_M$ with selected values of $\bar\gamma_E$ for the measured V2V channel. As observed for the previous two scenarios, the probability of SPSC is improved when $\bar\gamma_M$ \textgreater \ $\bar\gamma_E$.

%%%%%%%%%%%%%%%%%%%%%%%%%%%%%%%%%%%%%%%%%%%%%%%%%%%%%%%%%%%
\section{Conclusion}
%%%%%%%%%%%%%%%%%%%%%%%%%%%%%%%%%%%%%%%%%%%%%%%%%%%%%%%%%%%

%Fig. \ref{fig1_test} is a test. \figref{fig1_test} is also a test.% 

Novel analytical and closed-form expressions for the probability of SPSC and SOP$^L$ of the recently proposed $\kappa$-$\mu$ fading model have been presented. Specifically, the analytical expressions have been derived for positive, real, \textit{i.n.i.d.} $\kappa$-$\mu$ channel variables in the presence of an eavesdropper. We have also arrived at an exact closed form expression for the probability of SPSC for integer values of $\mu_M$ and $\mu_E$. Based on these results we have provided a useful insight into the behavior of SPSC and SOP$^L$ as a function of parameters \{$\kappa_M$, $\mu_M$, $\bar\gamma_M$\} and \{$\kappa_E$, $\mu_E$, $\bar\gamma_E$\}. The analytical and closed form expressions have been validated through reduction to known special cases. As the $\kappa$-$\mu$ fading model is a very general statistical model that includes many well-known distributions, the new equations derived in this paper will find use in characterizing the secrecy performance of several different fading channels (see Table \ref{Table:1}). Moreover, the results presented here will also find immediate application in the analysis of outage probability in wireless systems affected by co-channel interference and background noise, and the calculation of outage probability in interference-limited scenarios. Finally, we have illustrated the utility of the new formulations by investigating the probability of SPSC based on real channel measurements conducted for a diverse range of wireless applications such as cellular device-to-device, vehicle-to-vehicle and body centric fading channels. It is also worth highlighting that all of the expressions presented in this paper can be easily evaluated using functions available in mathematical software packages such as Mathematica and Matlab.

\section*{Acknowledgment}
%%%%%%%%%%%%%%%%%%%%%%%%%%%%%%%%%%%%%%%%%%%%%%%%%%%%%%%%%%%

This work was supported by the U.K. Royal Academy of Engineering, the Engineering and Physical Sciences Research Council under Grant References EP/H044191/1 and EP/L026074/1 and also by the Leverhulme Trust, UK through PLP-2011-061.

%%%%%%%%%%%%%%%%%%%%%%%%%%%%%%%%%%%%%%%%%%%%%%%%%%%%%%%%%%%
\appendices
\section{Proof of Equation $\eqref{eq:1b}$\label{app:A}} 
%%%%%%%%%%%%%%%%%%%%%%%%%%%%%%%%%%%%%%%%%%%%%%%%%%%%%%%%%%%
From $\eqref{eq:1a}$ we have,
\begin{equation}
\begin{aligned}[b]
{P_{out}}\left( {{R_S}} \right) &= {P_r}\left[ {{\gamma _M} \le {e^{R_S} }\left( {1 + {\gamma _E}} \right) - 1} \right]\\
 & = \mathop \smallint \limits_0^\infty  {f_{{\gamma _E}}}\left( {{\gamma _E}} \right)\left[ {\mathop \smallint \limits_0^{{e^{R_S}  }\left( {1 + {\gamma _E}} \right) - 1}  {f_{{\gamma _M}}}\left( {{\gamma _M}} \right)d{\gamma _M}} \right]d{\gamma _E}\\
&  = \mathop \smallint \limits_0^\infty  {f_{{\gamma _E}}}\left( {{\gamma _E}} \right)\left[ {{F_{{\gamma _M}}}\left( {{e^{R_S}  }\left( {1 + {\gamma _E}} \right) - 1} \right)} \right]d{\gamma _E}  \label{eq:19}
\end{aligned}
\end{equation}
Substituting $\eqref{eq:6}$ and $\eqref{eq:7}$ in $\eqref{eq:19}$ we obtain $\eqref{eq:1b}$.

\section{Proof of Proposition $\ref{prop:1a}$\label{app:B}} 
An analytical expression for $\eqref{eq:1d}$ can be derived by expressing the generalized Marcum $Q$-function and the modified Bessel function of the first kind according to~\cite[eq. 16]{sofotasios2014analytic} and~\cite{HTF} as, follows,
\begin{equation}
{Q_m}\left( {a,b} \right) = \mathop \sum \limits_{l = 0}^\infty  \frac{{{a^{2l}}{\rm{\Gamma }}\left( {m + l,\frac{{{b^2}}}{2}} \right)}}{{l!{\rm{\Gamma }}\left( {m + l} \right){2^l}{e^{\frac{{{a^2}}}{2}}}}}\label{eq:21}
\end{equation}
\begin{equation}
{I_v}(x) = \mathop \sum \limits_{k = 0}^\infty  \frac{{{{\left( {\frac{x}{2}} \right)}^{v + 2k}}}}{{k!{\rm{\Gamma }}\left( {v + k + 1} \right)}}\label{eq:22}
\end{equation}
where $\Gamma(\cdot)$ is the gamma function, $\Gamma(\cdot,\cdot)$ is the incomplete gamma function. By following the procedure proposed in~\cite[Lemma 1]{sofotasios2014analytic}, the following upper bound is obtained for the truncation error of the Marcum $Q$- function representation in $\eqref{eq:21}$,
\begin{align*}
{_t} \le \sum\limits_{l = 1}^{m - \frac{1}{2}} {\sum\limits_{k = 0}^{l - 1} {\frac{{{{\left( {- 1} \right)}^l}{b^k}{{\left( {l - k} \right)}_{l - 1}}\left[{1-{{\left({-1}\right)}^k}{e^{2ab}}} \right]}}{{k!\sqrt \pi  {2^{l - k - \frac{1}{2}}}{a^{2l - k - 1}}{e^{\frac{{{{\left( {a + b} \right)}^2}}}{2}}}}}} } + Q\left( {a + b} \right)+ Q\left( {b - a} \right)-\mathop \sum \limits_{l = 0}^p \frac{{{a^{2l}}{\rm{\Gamma }}\left( {m + l,\frac{{{b^2}}}{2}} \right)}}{{l!{\rm{\Gamma }}\left( {m + l} \right){2^l}{e^{\frac{{{a^2}}}{2}}}}}
\end{align*}
By substituting $\eqref{eq:21}$ and $\eqref{eq:22}$ in $\eqref{eq:1d}$ we obtain,
\begin{align}
{\mathrm{SOP}^L} &= \frac{{{\beta _E}^{\frac{{{\mu _E} + 1}}{2}}}}{{{\alpha _E}^{\frac{{{\mu _E} - 1}}{2}}{e^{{\alpha _E}}}}}\sum\limits_{k = 0}^\infty  {\frac{{{{\left( {\sqrt {{\alpha _E}{\beta _E}} } \right)}^{{\mu _E} - 1 + 2k}}}}{{k!{\rm{\Gamma }}\left( {{\mu _E} + k} \right)}}} \mathop \smallint \limits_0^\infty  {\gamma _E}^{{\mu _E} + k - 1}{e^{ - {\beta _E}{\gamma _E}}}d{\gamma _E} - \frac{{{\beta _E}^{\frac{{{\mu _E} + 1}}{2}}}}{{{\alpha _E}^{\frac{{{\mu _E} - 1}}{2}}{e^{{\alpha _E}}}}} \nonumber\\
&\times \sum\limits_{k = 0}^\infty  {\sum\limits_{l = 0}^\infty  {\frac{{{{\left( {\sqrt {{\alpha _E}{\beta _E}} } \right)}^{{\mu _E} - 1 + 2k}}{{\left( {2{\alpha _M}} \right)}^l}}}{{k!{\rm{\Gamma }}\left( {{\mu _E} + k} \right)l!{\rm{\Gamma }}\left( {{\mu _M} + l} \right){2^{l}}{e^{{\alpha _M}}}}}} }\mathop \smallint \limits_0^\infty  {\gamma _E}^{{\mu _E} + k - 1}{e^{ - {\beta _E}{\gamma _E}}}{\rm{\Gamma }}\left( {{\mu _M} + l,{\beta _M}{e^{{R_S}}}{\gamma _E}} \right)d{\gamma _E} \label{eq:1g} \end{align}
where $a=  1/\bar\gamma_M$,  $b=  1/\bar\gamma_E$, ${\beta _M} = ({\kappa _M} + 1)a{\mu _M}$, ${\beta _E} = ({\kappa _E} + 1)b{\mu _E}$, ${\alpha _M} = {\kappa _M}{\mu _M}$ and ${\alpha _E} = {\kappa _E}{\mu _E}$. 
Notably, the integrals in $\eqref{eq:1g}$ are identical to~\cite[eq. 3.381,4]{TofI} and~\cite[eq. 6.455]{TofI} given by,
\begin{align}
\mathop \smallint \limits_0^\infty  {x^{v - 1}}{e^{ - \mu x}}dx = \frac{1}{{{\mu ^v}}}{\rm{\Gamma }}\left( v \right)
~~~~~~~~~~~~~~~~~~~~~~~~~~~~~~~~~~~~~~~~~~~~~\left[ {Re~\mu  > 0,Re~{v} > 0} \right]
\label{eq:1h}
\end{align}
and
\begin{multline}
\mathop \smallint \limits_0^\infty  {x^{\mu  - 1}}{e^{ - \beta x}}{\rm{\Gamma }}\left( {v,\alpha x} \right)dx = \frac{{{\alpha ^v}{\rm{\Gamma }}\left( {\mu  + v} \right)}}{{\mu {{\left( {\alpha  + \beta } \right)}^{\mu  + v}}}} \ {_2}{F_1}\left( {1,\mu  + v;\mu  + 1;\frac{\beta }{{\alpha  + \beta }}} \right)\\
\left[ {Re\left( {\alpha  + \beta } \right) > 0,Re~\mu  > 0,Re\left( {\mu  + v} \right) > 0} \right]
\label{eq:1i}
\end{multline}
We then substitute $\eqref{eq:1h}$ and $\eqref{eq:1i}$ in $\eqref{eq:1g}$ to obtain $\eqref{eq:1e}$.

\section{Proof of Equation $\eqref{eq:13}$\label{app:C}} 

From $\eqref{eq:1f}$ we have,
\begin{equation}
\begin{aligned}[b]
{P_0} &= {P_r}\left( {{\gamma_M} > {\gamma_E}} \right) \\
 & = \mathop \smallint \limits_0^\infty  {f_{{\gamma _E}}}\left( {{\gamma _E}} \right)\left[ {\mathop \smallint \limits_{{\gamma _E}}^\infty  {f_{{\gamma _M}}}\left( {{\gamma _M}} \right)d{\gamma _M}} \right]d{\gamma _E} \\
&=\mathop \smallint \limits_0^\infty  {f_{{\gamma _E}}}\left( {{\gamma _E}} \right)\left[ {1 - {F_{{\gamma _M}}}\left( {{\gamma _E}} \right)} \right]d{\gamma _E}. \label{eq:1j}
\end{aligned}
\end{equation}
Substituting $\eqref{eq:6}$ and $\eqref{eq:7}$ in $\eqref{eq:1j}$ we obtain $\eqref{eq:13}$.

\section{Proof of Proposition $\ref{prop:1}$\label{app:D}} 
An analytical expression for the probability of SPSC is obtained by substituting $\eqref{eq:21}$ and $\eqref{eq:22}$ in $\eqref{eq:13}$ as follows, 
\begin{align}
{P_0} &= \frac{{{\beta _E}^{\frac{{{\mu _E} + 1}}{2}}}}{{{\alpha _E}^{\frac{{{\mu _E} - 1}}{2}}{e^{{\alpha _E}}}}}\sum\limits_{k = 0}^\infty  {\sum\limits_{l = 0}^\infty  {\frac{{{{\left( {\sqrt {{\alpha _E}{\beta _E}} } \right)}^{{\mu _E} - 1 + 2k}}{{\left( {2{\alpha _M}} \right)}^l}}}{{k!{\rm{\Gamma }}\left( {{\mu _E} + k} \right)l!{\rm{\Gamma }}\left( {{\mu _M} + l} \right){2^{l}}{e^{{\alpha _M}}}}}} }\nonumber
\\ &\times  \mathop \smallint \limits_0^\infty  {\gamma _E}^{{\mu _E} + k - 1}{e^{ - {\beta _E}{\gamma _E}}}  {\rm{\Gamma }}\left( {{\mu _M} + l,{\beta _M}{\gamma _E}} \right)d{\gamma _E}\label{eq:23}
\end{align}
Notably, the above integral is identical to~\cite[eq. 6.455]{TofI} given by,
\begin{multline}
\mathop \smallint \limits_0^\infty  {x^{\mu  - 1}}{e^{ - \beta x}}{\rm{\Gamma }}\left( {v,\alpha x} \right)dx = \frac{{{\alpha ^v}{\rm{\Gamma }}\left( {\mu  + v} \right)}}{{\mu {{\left( {\alpha  + \beta } \right)}^{\mu  + v}}}} \ {_2}{F_1}\left( {1,\mu  + v;\mu  + 1;\frac{\beta }{{\alpha  + \beta }}} \right)\\
\left[ {Re\left( {\alpha  + \beta } \right) > 0,Re~\mu  > 0,Re\left( {\mu  + v} \right) > 0} \right]
\label{eq:24}
\end{multline}
We then substitute $\eqref{eq:24}$ in $\eqref{eq:23}$ to obtain $\eqref{eq:14}$.

\section{Proof of Proposition $\ref{prop:2}$\label{app:E}} 
From~\cite[eq. 2.5]{price1964some},
\begin{equation}
{P_{\mu ,v}}\left( {A,B;r} \right) = {A^{ - \mu }}{B^{ - v}}\mathop \smallint \limits_0^\infty  {x^{\mu  + 1}}{e^{ - \left( {\frac{{{x^2} + {A^2}}}{2}} \right)}}{I_\mu }\left( {Ax} \right)dx \mathop \smallint \limits_0^{rx} {y^{v + 1}}{e^{ - \left( {\frac{{{y^2} + {B^2}}}{2}} \right)}}{I_v}\left( {By} \right)dy.
\label{eq:25}\end{equation}
Using the definition of the Marcum $Q$-function given in $\eqref{eq:4}$ in $\eqref{eq:25}$, we obtain,
\begin{equation}
{P_{\mu ,v}}\left( {A,B;r} \right) = 1 - {A^{ - \mu }}\mathop \smallint \limits_0^\infty  {x^{\mu  + 1}}{e^{ - \left( {\frac{{{x^2} + {A^2}}}{2}} \right)}}{I_\mu }\left( {Ax} \right){Q_{v + 1}}\left( {B,rx} \right)dx.
\label{eq:26}\end{equation}
Now letting $x = \sqrt {\frac{{2{\mu _E}\left( {1 + {\kappa _E}} \right){\gamma _E}}}{{\overline {{\gamma _E}} }}}$ and performing the necessary transformation of variables we obtain,
\begin{multline}
{P_{\mu ,v}}\left( {A,B;r} \right) = 1 - {A^{ - \mu }}{e^{ - \frac{{{A^2}}}{2}}}\mathop \smallint \limits_0^\infty  {2^{\frac{\mu }{2}}}{\left( {\frac{{{\mu _E}\left( {1 + {\kappa _E}} \right)}}{{\overline {{\gamma _E}} }}} \right)^{\frac{\mu }{2} + 1}}{\gamma _E}^{\frac{\mu }{2}}{e^{\frac{{ - {\mu _E}\left( {1 + {\kappa _E}} \right){\gamma _E}}}{{\overline {{\gamma _E}} }}}}\\
 \times {I_\mu }\left( {A\sqrt {\frac{{2{\mu _E}\left( {1 + {\kappa _E}} \right){\gamma _E}}}{{\overline {{\gamma _E}} }}} } \right){Q_{v + 1}}\left( {B,r\sqrt {\frac{{2{\mu _E}\left( {1 + {\kappa _E}} \right){\gamma _E}}}{{\overline {{\gamma _E}} }}} } \right)d{\gamma _E}.
\label{eq:27}\end{multline}
Comparing $\eqref{eq:27}$ with $\eqref{eq:13}$ and with the appropriate variable substitutions (see Proposition $\ref{prop:2}$), we obtain:
\begin{equation}
\begin{aligned}[b]
{P_{\mu ,v}}\left( {A,B;r} \right) =& 1 - \frac{{{\mu _E}{{\left( {1 + {\kappa _E}} \right)}^{\frac{{{\mu _E} + 1}}{2}}}}}{{{\kappa _E}^{\frac{{{\mu _E} - 1}}{2}}{{\overline {{\gamma _E}} }^{\frac{{{\mu _E} + 1}}{2}}}{e^{{\mu _E}{\kappa _E}}}}}\mathop \smallint \limits_0^\infty  {\gamma _E}^{\frac{{{\mu _E} - 1}}{2}}{e^{\frac{{ - {\mu _E}\left( {1 + {\kappa _E}} \right){\gamma _E}}}{{\overline {{\gamma _E}} }}}} \\
 & \!\!\!\!\!\times\!{I_{{\mu _E} - 1}}\!\left( {\!2{\mu _E}\sqrt {\frac{{{\kappa _E}\left( {1 + {\kappa _E}} \right){\gamma _E}}}{{\overline {{\gamma _E}} }}} } \right)
 \!{Q_{{\mu _M}}}\!\left(\! {\sqrt {2{\kappa _M}{\mu _M}} ,\sqrt {\frac{{2{\mu _M}\left( {1 + {\kappa _M}} \right){\gamma _E}}}{{\overline {{\gamma _M}} }}} } \right)d{\gamma _E}\\
 =& 1 - {P_0}. \label{eq:28}
\end{aligned}
\end{equation}
From ~\cite[eq. 3.16]{price1964some}, we have,
\begin{multline}
{P_{\mu ,v}}\left( {A,B;r} \right) = {P_{0,0}}\left( {A,B;r} \right) + exp\left( { - \frac{{{A^2}r + {B^2}{r^{ - 1}}}}{{2R}}} \right)\mathop \sum \limits_{m =  - \mu }^v {\left( {\frac{A}{{Br}}} \right)^m}{I_m}\left( {\frac{{AB}}{R}} \right)\\
 \times \left\{ {\mathop \sum \limits_{k = 1}^\mu  \left( {\begin{array}{*{20}{c}}
{v + k}\\
{k + m}
\end{array}} \right){r^{v - k + 1}}{R^{ - v - k - 1}} - \mathop \sum \limits_{j = 1}^v \left( {\begin{array}{*{20}{c}}
j\\
m
\end{array}} \right){r^{j - 1}}{R^{ - j - 1}}} \right\}.
\label{eq:29}\end{multline}
From~\cite[eq. 3.5]{price1964some}, we have,
\begin{equation}
{P_{0,0}}\left( {A,B;r} \right) = Q\left( {\frac{{Ar}}{{\sqrt {1 + {r^2}} }},\frac{B}{{\sqrt {1 + {r^2}} }}} \right) - {\left( {1 + {r^2}} \right)^{ - 1}}\exp \left[ { - \frac{{{A^2}{r^2} + {B^2}}}{{2\left( {1 + {r^2}} \right)}}} \right]{I_0}\left( {\frac{{ABr}}{{1 + {r^2}}}} \right).
\label{eq:30}
\end{equation}
Letting $\acute{P} = {P_{0,0}}\left( {A,B;r} \right)$ and combining $\eqref{eq:28}$, $\eqref{eq:29}$ and $\eqref{eq:30}$ we obtain $\eqref{eq:15}$.
Note that~\cite{price1964some} uses lower-case symbols (a, b) and we use upper-case symbols (A, B). This is because we define $a =  \frac{1}{\bar\gamma_M}$ and $b =  \frac{1}{\bar\gamma_E}$.

% \bibliographystyle{IEEEtran}
%  \begin{thebibliography}{99}
 
% \bibitem{ref01} 
% A. Ghosh, N. Mangalvedhe, R. Ratasuk, B. Mondal, M. Cudak, 
% E. Visotsky, T. A. Thomas, J. G. Andrews, P. Xia, H. S. Jo, H. S. Dhillon, and T. D. Novlan, 
% ``Heterogeneous cellular networks: From theory to practice,'' 
% \textit{IEEE Communications Magazine}, vol. 50, no. 6, pp.54-64, June 2012.

% \bibitem{ref01_ver2} 
% J. G. Andrews, 
% ``Seven ways that HetNets are a cellular paradigm shift,'' 
% \textit{IEEE Communications Magazine}, vol. 51, no. 3, pp.136-144, March 2013.

% \bibitem{ref02}
% M. Haenggi, 
% \textit{Stochastic Geometry for Wireless Networks}, Cambridge University Press, 2012.

% \bibitem{ref02_ver2}
% M. Haenggi, 
% ``On distances in uniformly random networks,''
% \textit{IEEE Transactions on Information Theory}, vol. 51, no. 10, pp.3584-3586, Oct. 2005.

% \bibitem{ref02_ver3}
% F. Baccelli and B. Blaszczyszyn, 
% ``Stochastic Geometry and Wireless Networks: Applications,''
% \textit{Foundations and Trends in Networking}, vol. 4, no. 1-2, pp. 1-312, 2009.

% \bibitem{ref10}
% H. S. Dhillon, R. K. Ganti, F. Baccelli and J. G. Andrews, 
% ``Modeling and Analysis of K-Tier Downlink Heterogeneous Cellular Networks,''
% \textit{IEEE Journal on Selected Areas in Communications}, vol. 30, no. 3, pp. 550-560, Apr. 2012.

% \bibitem{ref12}
% M. Haenggi, ``A Versatile Dependent Model for Heterogeneous Cellular Networks,'' 
% May 2013, available at http://arxiv.org/abs/1305.0947.

% \end{thebibliography} 

\nocite{*}
\bibliographystyle{IEEEtran}
\bibliography{IEEEabrv,ref}

%%%%%%%%%%%%%%%%%%%%%%%%%%%%%%%%%%%%%%%%%%%%%%%%%%%%%%%%%%%

% \begin{figure*}[!t]
%     \centering
%     \includegraphics[width=0.5\linewidth]{pointprocess_b.eps}
%     \caption{Stochastic dependencies amongst the point processes.}
%     \label{fig.pp}
% \end{figure*}

% \begin{figure*}[!t]
%     \centering
%     \includegraphics[width=0.5\linewidth, height =0.5\linewidth]{case1_b.eps}
%     \caption{Probability of joint occurrence versus the path loss exponent, illustrating the impact of the spatial correlation.}
%     \label{fig.spatial}
% \end{figure*}

% \begin{figure*}[!t]
%     \centering
%     \includegraphics[width=0.5\linewidth, height =0.5\linewidth]{case2_b.eps}
%     \caption{Mean local delay versus the ALOHA parameter $p$, illustrating the impact of the temporal correlation.}
%     \label{fig.temporal}
% \end{figure*}

% \begin{figure*}[!t]
%     \centering
%     \includegraphics[width=0.5\linewidth, height =0.5\linewidth]{case3_b.eps}
%     \caption{End-to-end outage probability versus the relay location, illustrating the impact of the spatial-temporal correlation.}
%     \label{fig.spatial-temporal}
% \end{figure*}

\end{document}